# STATICS OF LOOSE TRIANGULAR EMBANKMENT UNDER NADAI'S SAND HILL ANALOGY

Thirapong Pipatpongsa

*Global Scientific Information and Computing Center, Tokyo Institute of Technology, 2-12-1 O-okayama, Meguro-ku, Tokyo 152-8550, Japan*

Tel: +81-3-5734-2121

Fax: +81-3-5734-3276

pthira@gsic.titech.ac.jp

Sokbil Heng

*Department of International Development Engineering, Tokyo Institute of Technology, 2-12-1 O-okayama, Meguro-ku, Tokyo 152-8550, Japan*

Tel: +81-3-5734-2121

Fax: +81-3-5734-3276

hengsokbil@ide.titech.ac.jp

Atsushi Iizuka

*Research Center for Urban Safety and Security, Kobe University, 1-1 Rokkodai-cho, Nada-ku, Kobe 657-8501, Japan*

Tel: +81-78-803-6437

Fax: +81-78-803-6394

iizuka@kobe-u.ac.jp

Hideki Ohta

*Research and Development Initiative, Chuo University, 1-13-27 Kasuga, Bunkyo-ku, Tokyo 112-8551, Japan*

Tel: +81-3-3817-1637

Fax: +81-3-3817-1606

ohta@tamacc.chuo-u.ac.jp

Corresponding author: Thirapong Pipatpongsa, E-mail: pthira@gsic.titech.ac.jp

**ABSTRACT**

In structural mechanics, Nadai's sand hill analogy is the interpretation of an ultimate torque applied to a given structural member with a magnitude that is analogously twice the volume of stable sand heap which can be accommodated on a transverse cross-section basis. Nadai's analogy is accompanied by his observation of a loose triangular embankment, based on the fact that gravitating loose earth is stable if inclined just under the angle of repose. However, Nadai's analysis of stress distribution in a planar sand heap was found to be inaccurate because the total pressure obtained from Nadai's solution is greater than the self-weight calculated from the heap geometry. This raises a question about the validity of his observation in relation to the analogy. To confirm his criterion, this article presents and corrects the error found in Nadai's solution by analyzing a radially symmetric stress field for a wedge-shaped sand heap with the purpose of satisfying both force balance and Nadai's closure. The fundamental equation was obtained by letting the friction state vary as a function of angular position and deduce it under the constraint that the principal stress orientation obeys Nadai's closure. The theoretical solution sufficiently agreed with the past experimental measurements.

## 1. Introduction

Rapid development in the subject of plasticity appeared with the publication of the classical books of Arpad L. Nádai (Nadai, 1963). His works are considered as significant contributions to the field of engineering materials, particularly to the areas of structural mechanics and geo-mechanics. Nadai's sand hill or sand heap analogy is widely known as the graphical interpretation of a fully plastic condition progressing throughout a twisted member. The stress function solution can be visually depicted as a surface of sand heap which is poured upon a horizontal plate shaped in a cross-section. This analogy is based on the fact that a slope of dry sand has a natural angle of repose that is slightly greater than the angle of internal friction. A heap of sand tries to keep a constant slope everywhere because the sand particles upslope will cascade down once the slope exceeds this limit. Therefore, if sand is added slowly and continually, a heap will grow and reach a unique steady state, with the shape remaining unchanged with the appearance of the peak, edges or ridge line. Nadai observed that by replacing a frictional coefficient to ultimate shear stress, the ultimate torque applied to any given structural member is analogously twice the volume of stable sand heap growing on that cross-section. An outline of his analogy is elaborated on in **Fig.1** and **Fig.2**. A detailed explanation can be found in the first volume of his monograph and other relevant textbooks in structural mechanics and plasticity theory (Calladine, 2000; Kachanov, 1971; Richards, 2000).

Nadai's achievements in geo-mechanics began with the second volume of his monograph. It is evident that Nadai also tried to investigate stress distribution in a planar sand heap under plane-strain condition. In accordance with his sand hill analogy, Nadai pointed out that gravitating loose earth is stable if inclined at an angle just lower than the angle of repose. Any small lump of sand added to the surface of a heap will cause a flow confined only to the surface, so the deeper sand lying below the surface remains stable and immobilized. Let us regard his observation of sand heap as Nadai's sand heap criterion. Experiments on various topographies of steady sand heaps verifying Nadai's sand heap criterion can be found in Pauli and Gioia (2007). Some basic experiments carried out by the authors are also demonstrated in **Fig.3** and **Fig.4** to verify that the final shape of the sand heap is independent of the sand deposition method used.

To obtain the stress solution, Nadai considered a rigid-elastic body of a long sand heap whose symmetrical surfaces are stabilized just under the Coulomb failure criterion with zero cohesion. Nadai noticed that an infinitesimal interface of zero traction resisted by friction along the sliding plane can provide a major compressive direction in accordance with the Coulomb failure criterion of pure friction material, which is angled precisely at the middle of the slope of a sliding plane and the direction of gravity. Moreover, the condition of zero shear stress along the central plane clearly reveals that the major compressive direction under the ridge is parallel to the gravity direction. According to two known limits of direction of major compressive stress, the assumed linear relation between the angle of major compressive stress and the polar angle measured from the ridge was introduced. Finally, the states of stress satisfying the equilibrium configuration under self-weight loading were formulated. However, it was found later that his solution contains an error of inappropriately associating vertical stress directly under the ridge of the sand heap equal to the height of the heap times the bulk unit weight of sand. By virtue of partial support from frictional resistance among sand particles, it is understandable that the weight of a thin column of sand under the apex cannot be wholly transferred to the base. In fact, none of the experimental results in wedge-shaped sand heap confirm that the basal pressure in the middle of the heap represents a full geostatic pressure, e.g. Hummel and Finnan (1921), Trollope (1956), Lee & Herinton (1971), Vanel, et al. (1999) and Wiesner (2000).

Nadai's solution was briefly reviewed by Marais (1969) in a study of stresses in heaps of cohesionless sand. Through integration of vertical pressure exerted on the base, one could find that the vertical thrust given by Nadai's solution is greater than the weight of the sand heap, hence violating the equilibrium condition in a vertical direction. This error reflects the incorrect boundary condition set as a result of erroneous intuition on the central pressure. As a result, Marais made a correction to Nadai's solution by equating the vertical thrust acting on the base to the weight of the sand heap in an attempt to replace the imposed boundary condition of the ridge's full geo-static pressure to a whole weight-balanced constraint condition. Nevertheless, Marais considered his correction to Nadai's solution as being an estimation to Sokolovskii (1965)'s limiting stress solution on the planar heap, in which states of stress saturate the Coulomb yield criterion everywhere. This conclusion on limit equilibrium in a sand heap contradicts the original concept of Nadai's sand heap criterion because only states of stress along the slope surface yields, but those bound inside do not.

Because a sand heap at rest with a slope lower than the angle of repose is solid-like in behavior, other varying assumptions (Cantelaube and Goddard, 1997; Didwania, et al., 2000; Wittmer, et al., 1997; Wittmer, et al., 1996) suggest an elastic region theoretically exists below the yielding limit. In contrast with limiting equilibrium in a sand heap, state of stress does not reach a fully mobilized state of active condition everywhere. Rather, it does so only partially in the outer crust, leaving the admissible state below-yielding criterion in the inner core. However, their assumptions require that the outer plastic region at yielding limit begins to develop from deeper within the sliding surface, e.g. as much as one-third of bulk volume, based on the closure of fixed principal axes which is applied to a sand heap with an angle of repose $33^{o}$ (Wittmer, et al., 1997; Wittmer, et al., 1996).

The developed yielding region is inconsistent with experimental observations (Jaeger, et al., 1996). The flow of dry sand on a tilted slope just above the angle of repose is clearly different from that of liquid or slurry because the flow occurs only on a boundary layer at the heap's surface, with no motion in the bulk at all. The particles deeper within the heap do not participate in the motion when other particles along the sliding surface start to flow. This observation perfectly agrees with Nadai's sand heap criterion.

The authors found that both Nadai's solution and Marais's correction satisfy the equilibrium condition in the horizontal direction but violate the differential equations of equilibrium in the vertical direction. These unsatisfactory results led to a theoretical concern as to whether the criterion of Nadai's sand heap really does exist. This article aims to clarify the misconception in earlier derivations of stress distribution in a planar sand heap inclined at the angle of repose. The sources of error are reviewed and the exact formulation of stress distribution is presented. The rigorous formulation realizes the theoretical correction that completely satisfies the criteria of Nadai's sand heap and does not violate the equilibrium and boundary conditions. Finally, the stress distributions are validated by some published experimental results on prismatic wedges of poured sand.

Hence, the result of this study can mathematically justify Nadai's sand heap criterion and reaffirms the basic concept of Nadai's sand hill analogy. The method may be extended to include a conical shape or arbitrary domains of the base in succeeding research.

## 2. Theory of planar Heap

### 2.1. Stability in a sand heap

Compression is considered as positive throughout this study. An ideal heap, composed of perfectly loose and uniformly distributed grains of dry sand, is assumed. The component of shear stress $\tau_n$ and normal stress $\sigma_n$ in any orientation of the section must satisfy Coulomb's law of friction, where subscript $n$ refers to the normal direction of the section. Therefore, for cohesionless materials, the limiting states of stress are expressed by the following inequalities, either in terms of stress components or principal stresses, where $\phi$ denotes the angle of friction, $\sigma_1$ and $\sigma_3$ represents the major and minor principal stresses, respectively. As soon as the slip starts, the inequality sign will be replaced by the equality sign.

$$|\tau_n|/\sigma_n \le \tan\phi \tag{1}$$

$$(\sigma_1 - \sigma_3)/(\sigma_1 + \sigma_3) \le \sin\phi \tag{2}$$

According to Nadai's sand heap criterion, only the state of stress along the sliding plane is mobilized by friction. Let us consider the state of stress by means of the Mohr stress circle drawn in **Fig.5,** where $\sigma_f$ and $\tau_f$ denote the normal stress and positive shear stress at the particular orientation passing the failure envelope. Therefore, the inequality sign in Eq.(1) is replaced with the equality sign and a subscript $n$ is specified to $f$ as shown in Eq.(3) to specify the failure condition. For a loose deposit of granular wedges, the angle of internal friction is commonly taken to be equivalent to the angle of repose. Hence, $(\sigma_f,\tau_f)$ itself is inclined parallel to the slope of the sliding plane. The major and minor principal stress $\sigma_1$ and $\sigma_3$, where $\sigma_1 \ge \sigma_3$, can be obtained and remarked as $\sigma_{1f}$ and $\sigma_{3f}$ for the mobilized state in the same circle passing $(\sigma_f,\tau_f)$. Therefore, the relation given by Eq.(3) can be alternatively presented in Eq.(4) using the states of principal stress.

$$\tau_f/\sigma_f = \tan\phi \tag{3}$$

$$(\sigma_{1f} - \sigma_{3f})/(\sigma_{1f} + \sigma_{3f}) = \sin\phi \tag{4}$$

Furthermore, the states of stress on various planes can be described by taking $(\sigma_f,\tau_f)$ as the pole of the Mohr circle shown in **Fig.5**. Correspondingly, the direction of the major principal stress along the sliding surface upon failure can be obtained and represented by the angle $\omega_f$. By the geometry of a triangle passing $(\sigma_f,\tau_f)$, $(\sigma_{3f},0)$ and $(\sigma_{1f},0)$, the relation of $\omega_f$ with $\phi$ is obtained thru Eq.(5). The angle bisects the angle between the vertical and the sliding surface.

$$\omega_f = \frac{\pi}{4} - \frac{\phi}{2} \tag{5}$$

## 2.2. States of stress in a planar sand heap

In **Fig.2**, a typical geometry of a sand heap of height $h$ deposited on a rectangular base with dimensions $a \times b$ is shown. Let us now consider a long sand heap whose dimension $a$ is much greater than $b$. The normal $n$ would turn around the direction of the intermediate stress $\sigma_2$ defined by the longitudinal direction in a sand heap. Herein, the quantity of $\sigma_2$ is of no interest, and the problem can be reduced to a plane strain condition. In **Fig.6**, the geometry of a planar sand heap in plane $xz$ (a transverse cross-section also shown by the shaded area in **Fig.2**) is bound by the symmetrical slope surfaces and the horizontal plane of the base, where axis $z$ and axis $x$ denote the direction of gravity and horizontal direction, respectively, as shown in **Fig.6**.

Radial distance $r$ is measured from the origin of axes marked on the ridge, while the angular coordinate $\theta$ is taken as positive with the anti-clockwise direction indicated in **Figs. 6-7**. In accordance with the polar coordinate system (see **Fig.7**) where $\gamma$ is the specific weight of the sand heap, the states of stress of a sand heap loaded by its own weight must not violate the following conditions of equilibrium in plane polar coordinate $(r,\theta)$ denoting $\sigma_r$ and $\sigma_\theta$ to be radial and tangential normal stresses, and $\tau_{r\theta}$ to be shear stress.

$$\partial_r \sigma_r + \frac{1}{r}\partial_\theta \tau_{r\theta} + \frac{\sigma_r - \sigma_\theta}{r} = \gamma \cos\theta \tag{6}$$

$$\partial_r \tau_{r\theta} + \frac{1}{r}\partial_\theta \sigma_\theta + \frac{2\tau_{r\theta}}{r} = -\gamma \sin\theta \tag{7}$$

Due to the heap's symmetry, the half-width heap in the positive angular range is considered for simplicity. Therefore, only the solution for the right side of the sand heap depicted in **Fig.6** is studied. The solution for the left side is obtained by mirroring the solution for the right side once it has been determined. Orientations of the principal compressive stress $\omega$ and $\psi$ are measured in an anti-clockwise direction from the $z$-axis and $r$-axis for each referenced coordinate system. Based on the axis transformation, a geometrical relation between $\omega$ and $\psi$ as well as between $r$ and $z$ were found as follows.

$$\psi = \omega - \theta \tag{8}$$

$$r = z/\cos\theta \tag{9}$$

Three components of stress $\sigma_r$, $\sigma_\theta$ and and $\tau_{r\theta}$ can be expressed in the form of stress invariant $p$ and deviatoric stress $q$ as described below:

$$\sigma_r = p(1+\beta\cos 2\psi) \tag{10}$$

$$\sigma_\theta = p(1-\beta\cos 2\psi) \tag{11}$$

$$\tau_{r\theta} = p\beta\sin 2\psi \tag{12}$$

where $p$ and $q$ implies the center and radius of the Mohr stress circle, while a frictional variable $\beta$ represents a ratio of $q/p$.

$$p = \frac{\sigma_1 + \sigma_3}{2} = \frac{\sigma_r + \sigma_\theta}{2} \tag{13}$$

$$q = \frac{\sigma_1 - \sigma_3}{2} = \sqrt{\left(\frac{\sigma_r - \sigma_\theta}{2}\right)^2 + \tau_{r\theta}^2} \tag{14}$$

$$\beta = \frac{\sigma_1 - \sigma_3}{\sigma_1 + \sigma_3} = q/p \tag{15}$$

Based on the rectangular coordinate system, stress components can be transformed to $\sigma_x$, $\sigma_z$ and $\tau_{xz}$ using the following expressions.

$$\sigma_x = p(1 - \beta \cos 2\omega) \tag{16}$$

$$\sigma_z = p(1 + \beta \cos 2\omega) \tag{17}$$

$$\tau_{xz} = p\beta \sin 2\omega \tag{18}$$

The relations between $\psi$ and the stress components in the polar coordinate system as well as between $\omega$ and stress components in the rectangular coordinate system are described below:

$$\tan 2\psi = 2|\tau_{r\theta}|/(\sigma_r - \sigma_\theta) \tag{19}$$

$$\tan 2\omega = 2|\tau_{xz}|/(\sigma_z - \sigma_x) \tag{20}$$

Despite different notations in the variables, referenced axes and sign convention, detailed explanations on the mathematical background used in this section can be found in many references (Cantelaube and Goddard, 1997; Cates, et al., 1998; Didwania, et al., 2000; Marais, 1969; Nadai, 1963; Sokolovskii, 1965; Wittmer, et al., 1997).

## 3. Limiting stress solutions

### 3.1. Sokolovskii's solution

By ignoring the scale length effect, it is possible to find the similarity solution in which the stress states proportionally increase with depth. Nadai (1963) as well as Sokolovskii (1965) guessed a form of in-plane mean stress $p$ to be a linear function of radius $r$. Many problems on the equilibrium of a sand heap possessing self-weight are also studied on the assumption that all the stress components $\sigma_r$, $\sigma_\theta$ and $\tau_{r\theta}$ are proportional to depth z or radius $r$ (Cantelaube and Goddard, 1997; Cates, et al., 1998;

Didwania, et al., 2000; Marais, 1969; Michalowski and Park, 2004; Wittmer, et al., 1997; Wittmer, et al., 1996). Therefore, a scaling factor of $p$ to geo-static pressure $\gamma r$ is defined to a non-dimensional scaled stress variable $\chi$ which is suggested to be a function of angle $\theta$ alone. This form is compatible with an angle of major principal stress $\psi$ which is dependent solely on $\theta$.

$$p = \gamma r \chi \tag{21}$$

This postulation is similar to a notion of radial stress field used by Sokolovskii; i.e. $\chi=\chi(\theta)$, $\psi=\psi(\theta)$. The substitution of stress components in Eqs.(10)-(12) with the equilibrium equations using Eq.(21) can formulate the following set of 1st order coupled differential equations.

$$\chi' = \frac{d\chi}{d\theta} = \frac{-\sin(2\psi+\theta)+\chi\sin 2\psi}{\cos 2\psi - \beta} \tag{22}$$

$$\psi' = \frac{d\psi}{d\theta} = \frac{\cos\theta - \beta\cos(2\psi+\theta)-(1-\beta^2)\chi}{2\beta\chi(\cos 2\psi - \beta)} - 1 \tag{23}$$

Sokolovskii considered a limiting equilibrium taken from Eq.(4) by imposing a constant gradient $\beta=\sin\phi$ throughout a sand heap. Two coupled systems of ordinary differential equations can be solved using the boundary conditions, consisting of a symmetry condition along the central plane where shear stress $\tau_{r\theta}$ is zero, and a stress-free condition along the sliding plane where $\sigma_r$ , $\sigma_\theta$ and $\tau_{r\theta}$ vanish to zero. Typically, the following boundary conditions for $\psi$ and $\chi$ are employed where subscript $c$ and $f$ denotes the location along the central plane and failure plane, respectively:

$$\psi_c = \psi|_{\theta=\theta_c} = 0 \quad \text{where} \quad \theta_c = 0 \tag{24}$$

$$\chi_f = \chi|_{\theta=\theta_f} = 0 \quad \text{where} \quad \theta_f = \pi/2 - \phi \tag{25}$$

According to recent research on the limit equilibrium of a granular heap (Cox, et al., 2008), the exact solution for the system of Eqs.(22)-(23) has yet to be found, with the exception of $\beta=1$ or $\phi=\pi/2$. Therefore, the numerical solution is generally undertaken to obtain Sokolovskii's solution.

### 3.2. Nadai's solution

Nadai considered the range of $\omega$ between $\omega_c \le \omega \le \omega_f$ along the central plane to the sliding plane. The minimum magnitude of $\omega$ is at the center according to Eqs.(8) and (24), while the maximum magnitude of $\omega$ is $\omega_f$ at the slope surface according to Eq.(5).

$$\omega_c = \omega|_{\theta=\theta_c} = 0 \tag{26}$$

$$\omega_f = \omega|_{\theta=\theta_f} = \pi/4 - \phi/2 \tag{27}$$

Nadai assumed $\omega$ to be a linear function of $\theta$ based on two extreme conditions given by Eqs.(26) and (27):

$$\omega = \frac{\omega_f - \omega_c}{\theta_f - \theta_c}(\theta - \theta_c) + \omega_c = \frac{\theta}{2} \tag{28}$$

The description of $\psi$, as well as derivatives, can be simply related to $\theta$ using the relation of $\psi$ and $\omega$ set in Eq.(8). The minus sign presented in Eq.(29) implies that $\psi$ tilts in a clockwise direction from the $r$-axis under Nadai's criterion.

$$\psi = -\theta/2 \tag{29}$$

$$\psi' = d\psi/d\theta = -1/2 \tag{30}$$

Substituting Eq.(29) to Eq.(22) using $\beta$=sin$\phi$ results in the differential equation for the function of $\chi$.

$$\chi' = \frac{\sin\theta}{\sin\phi - \cos\theta}\chi \tag{31}$$

Correspondingly, $\chi$ is found to be equal to the following equation with an integrating constant $c$, satisfying the stress-free condition along the sliding plane; i.e. for $\theta=\theta_f$, all stress components are zero, as indicated by Eq.(25):

$$\chi = c(\cos\theta - \sin\phi) \tag{32}$$

The vertical stress $\sigma_z$ underneath a heap with an arbitrary height $z$ above any horizontal plane can be expressed by substituting for $\chi$ from Eq.(32), $\omega$ from Eq.(28) to Eq.(17), $r$ from Eq.(9), using $\beta$=sin$\phi$.

$$\sigma_z = \gamma z c(\cos\theta - \sin\phi)(1 + \sin\phi\cos\theta)/\cos\theta \tag{33}$$

In an attempt to solve a constant $c$, the basal pressure directly under the ridge of the sand heap along the vertical symmetry axis, which coincides with the major compressive axis, is considered. However, Nadai incorrectly imposed that the pressure due to the weight of a thin column of sand above the base is $\gamma h$. This condition requires $\sigma_z=\gamma h$ under the ridge, therefore upon substituting $\theta=0$ to Eq.(33), a constant $c=-1/\cos^2\phi$ is obtained. Accordingly, $\chi$ in Eq.(32) can be expressed as follows:

$$\chi = \frac{\cos\theta - \sin\phi}{\cos^2\phi} \tag{34}$$

With $\beta=\sin\phi$ and Eq.(23) taking $\psi$ and $\psi'$ from Eqs.(29) and (30), we found that the alternative $\chi$ can be directly obtained without invoking the boundary condition:

$$\chi = \frac{\cos\theta - \sin\phi}{\cos 2\phi + \sin\phi\cos\theta} \tag{35}$$

One could verify Eqs.(34) and (35) and find that both solutions are not correct; i.e. substitution of Eq.(34) to Eq.(23) using Eq. (29) gives $\psi'=-1$ instead of $\psi'=-1/2$, as obtained in Eq.(30), while Eq.(35) is not parallel with Eq.(31). This error was later recognized and corrected by Marais (1969) and will be explained in the next section.

### 3.3. Marais's Solution

Marais (1969) purported that Eq.(29) replaces Eq.(23) when solved by approximation, so the result of Eqs.(34)-(35) can be ignored. He reconsidered Eq.(33), which was previously solved by Nadai. In order to satisfy force equilibrium in the vertical direction, Marais managed to refine the unknown integration constant $c$ by equating the weight of the sand heap to the vertical thrust. Herein, the weight of half-width wedge $W$ is determined from the volumetric integration over a half-width $h\cot\phi$.

$$W = \int_0^h \int_0^{h\cot\phi} \gamma\, dx\, dz = \frac{1}{2}\gamma h^2 \cot\phi \tag{36}$$

The vertical thrust $P$ acting on the half-width base is determined by integrating Eq.(33) over the base of the half-width wedge, taking $z=h$. Since $x=z\tan\theta$, $dx=(z/\cos^2\theta)d\theta$ can be employed as an integrating variable. The expression of $P$ can be obtained:

$$P = \int_0^{h\cot\phi} \sigma_z\, dx = \int_0^{\pi/2-\phi} \sigma_z \frac{h}{\cos^2\theta} d\theta = \frac{c\cot\phi}{2}\gamma h^2 \left( \cos 2\phi + \frac{\sin^2\phi}{\cos\phi} \ln\left( \frac{1+\cos\phi}{\sin\phi} \right) \right) \tag{37}$$

An integral constant $c$ is determined from the equilibrium of weight $W$ and vertical thrust $P$. Because shear stress vanishes along the central plane, there is no shearing support to the half-width heap along the separating symmetrical plane. Therefore, balance of force in a half-width heap is also taken to be the same as in a full-width heap. Equating $W$=$P$ for a half-width heap or 2$W$=2$P$ for a full-width heap results in the following expression of $c$:

$$c = \left( \cos 2\phi + \frac{\sin^2 \phi}{\cos \phi} \ln\left( \frac{1+\cos\phi}{\sin\phi} \right) \right)^{-1} \tag{38}$$

Then substituting $c$ to Eq.(32) corrects Nadai's original expression of $\chi$ to:

$$\chi = \frac{\cos\theta - \sin\phi}{\cos 2\phi + \frac{\sin^2 \phi}{\cos \phi} \ln\left( \frac{1+\cos\phi}{\sin\phi} \right)} \tag{39}$$

Despite the correction made, Eq.(39) does not satisfy Eq.(30) using Eq.(23), indicating a clear violation of the equilibrium condition. This perplexing problem of an error from an unknown source remained unsolved until the admissible stress theory developed by Michalowski and Park (Michalowski and Park, 2004) emerged.

## 4. Below-limiting stress solutions

### 4.1. Michalowski and Park's solution

The solutions of three stress components require three independent equations. Provided that one condition of principal axis trajectory is given by Eq.(19), two variables must involve two equilibrium equations stated in Eqs.(6)-(7). In the previous studies, we recognized that $\psi$ is treated as a known principal axis trajectory and $\chi$ was left as a single variable, in relation to the problem. Therefore, the source of the error is identified as one missing variable and the correction can be realized by considering Coulomb's frictional parameter $\beta$ as another variable rather than as a constant. The framework set forth in Michalowski and Park's admissible stress below-yielding criteria encompasses the method for solving the stress equations in a way similar to the limit equilibrium. Therefore, the inequality of friction law using a constant $\beta$ is replaced by the equality using a variable $\beta$. In our solution, we can associate $\beta$ with $\theta$ by taking $\beta=\beta(\theta)$. So the variable $\beta$, in terms of the angular position, allows us to address the admissible states of stress satisfying Eq.(2) where $\beta \le \sin\phi$ is expected. Accordingly, Eqs.(22)-(23) are modified by reconciling Eqs.(6)-(7) to the following set of differential equations, using Eqs.(10)-(12), Eq.(21) and $\beta'=d\beta/d\theta$:

$$\chi' = \frac{-\sin(2\psi+\theta)+(\sin 2\psi+\beta')\chi}{\cos 2\psi-\beta} \tag{40}$$

$$\psi' = \frac{\cos\theta-\beta\cos(2\psi+\theta)}{2\beta\chi(\cos 2\psi-\beta)} - \frac{1-\beta^2+\beta'\sin 2\psi}{2\beta(\cos 2\psi-\beta)} - 1 \tag{41}$$

It is clear that the above equations will be deduced to Eqs.(22)-(23) if $\beta'=0$. Equations (40)-(41) are equivalent to those derived by Michalowski and Park, in which their original forms are expressed in terms of $\phi=\phi(\theta)$.

The solution of below-limiting stress under the closure of polarized principal axes in Nadai's criterion is explained in the next section. Because the description of $\psi=\psi(\theta)$ is given, we shall rearrange Eqs.(40)-(41) to the following general set of unknown expressions relevant to $\chi$ and $\beta$ in our study:

$$\chi' = \frac{\cos(2\psi+\theta)-\left((3+2\psi')\beta+\cos 2\psi\right)\chi}{\sin 2\psi} \tag{42}$$

$$\beta' = \frac{\cos\theta-\beta\cos(2\psi+\theta)+\left((3+2\psi')\beta^2-2(1+\psi')\beta\cos 2\psi-1\right)\chi}{\chi\sin 2\psi} \tag{43}$$

### 4.2. Fundamental equation under Nadai's criterion

Substituting $\psi$ from Eq.(29) to Eqs.(42) and (43), we obtain a particular set of 1st order differential equations:

$$\chi' = -\frac{1-(\cos\theta+2\beta)\chi}{\sin\theta} \tag{44}$$

$$\beta' = -\frac{\cos\theta-\beta+\left(2\beta^2-\beta\cos\theta-1\right)\chi}{\chi\sin\theta} \tag{45}$$

These coupled derivative equations can be decoupled by taking more orders of derivatives until the substantial form is found. Further derivatives of Eq.(44) with $\theta$ obtains a form of $\chi''$ coupled with $\beta'$ and $\beta$:

$$\chi''=\frac{d^2\chi}{d\theta^2}=\frac{d\chi'}{d\theta}=\frac{\cos\theta+\left(2\beta+\cos\theta\right)\chi'\sin\theta-\left(1-2\beta'\sin\theta+2\beta\cos\theta\right)\chi}{\sin^2\theta} \tag{46}$$

In order to remove $\beta$ from Eq.(45) and Eq.(46), the expression for $\beta$ is simply rearranged from Eq.(44):

$$\beta=\frac{1-\chi\cos\theta+\chi'\sin\theta}{2\chi} \tag{47}$$

Substituting Eq.(47) to Eq.(45) achieves a form of $\beta'$ which is freed from $\beta$:

$$\beta'=\sin\theta+\left(3\cos\theta-\frac{1+\chi'\sin\theta}{\chi}\right)\frac{\chi'}{2\chi} \tag{48}$$

Substitutions of $\beta$ from Eq.(47) and $\beta'$ from Eq.(48) to Eq.(46) obtain a substantial form of $\chi''$ which is completely decoupled from $\beta$. Subsequently, the fundamental equation under Nadai's criterion can be arranged in a homogeneous second-order differential equation as follows:

$$\chi''-\left(2\cot\theta\right)\chi'-\chi=0 \tag{49}$$

This equation can be verified with those listed in Appendix A, where a fundamental equation in general formulation based on Eqs.(42) and (43) is derived using chain rule differentiation.

### 4.3. Solution under Nadai's criterion

Due to the characteristics of the differential equation shown in Eq.(49), the form of Eq.(49) can be reduced to the following second-order differential equation:

$$\cos\theta\left(\frac{\chi}{\cos\theta}\right)''-\frac{2}{\sin\theta}\left(\frac{\chi}{\cos\theta}\right)'=0 \tag{50}$$

where

$$\left(\frac{\chi}{\cos\theta}\right)'=\frac{\partial}{\partial\theta}\left(\frac{\chi}{\cos\theta}\right)=\frac{\chi'}{\cos\theta}+\frac{\sin\theta}{\cos^2\theta}\chi \tag{51}$$

$$\left(\frac{\chi}{\cos\theta}\right)''=\frac{\partial^2}{\partial\theta^2}\left(\frac{\chi}{\cos\theta}\right)=\frac{1}{\cos\theta}\chi''+\frac{2\sin\theta}{\cos^2\theta}\chi'+\frac{2}{\cos^3\theta}\chi-\frac{1}{\cos\theta}\chi \tag{52}$$

Let us replace $\chi$ in Eq.(50) by the following term, where $\lambda=\lambda(\theta)$ is a differentiable function of $\theta$. Therefore, $\chi$ is a linearly independent variable of $\cos\theta$.

$$\chi = \lambda \cos\theta \tag{53}$$

Substituting $\chi$ from Eq.(53), we can transform Eq.(50) to the second-order differential equation of variable $\lambda$ as expressed in terms of $\lambda'$ and $\lambda''$.

$$\lambda'' - \frac{2\lambda'}{\sin\theta\cos\theta} = 0 \tag{54}$$

Equation (54) can be solved using the order reduction technique. The solution in terms of the 1st order differential equation is found to be equal to the following equation with an integral constant $c_2$:

$$\lambda' = c_2 \frac{1-\cos 2\theta}{1+\cos 2\theta} \tag{55}$$

The following equation with an additional integral constant $c_1$ is found to be the solution of $\lambda$.

$$\lambda = c_1 + c_2\left(\tan\theta - \theta\right) \tag{56}$$

Substitution of Eq.(56) to Eq.(53) obtains the fundamental solution under Nadai's criterion.

$$\chi = \cos\theta\left(c_1 + c_2\left(\tan\theta - \theta\right)\right) \tag{57}$$

Two integrating constants are solved by two boundary conditions. Along the sliding surface, where $\theta=\theta_f$, the values of $\chi$ and $\chi'$ can be found. Because states of stress are zero along the sliding surface, therefore $\chi_f=0$ from Eq.(25) is taken as the primary boundary condition. The supplementary boundary condition is obtained from $\chi'_f$ in Eq.(44) by substituting $\theta=\theta_f$ and $\chi=\chi_f=0$ by virtue of Eq.(25).

$$\chi'_f = \chi'\big|_{\theta=\theta_f} = -1/\cos\phi \tag{58}$$

Consequently, $c_1$ and $c_2$ subject to the boundary conditions obtained from Eq.(25) and Eq.(58) can be solved.

$$c_1 = \frac{\cos\phi - \left(\frac{\pi}{2} - \phi\right)\sin\phi}{\cos^3\phi} \tag{59}$$

$$c_2 = -\frac{\sin\phi}{\cos^3\phi} \tag{60}$$

Finally, the particular solution of $\chi$ is derived in a closed form. Also, the expression of $\beta$ is obtained by Eq. (47) based on the derived expression of $\chi$.

$$\chi = \left(1 - \left(\frac{\pi}{2} - \phi - \theta + \tan\theta\right)\tan\phi\right)\frac{\cos\theta}{\cos^2\phi} \tag{61}$$

$$\beta = \frac{\left(\frac{\pi}{2} - \phi - \theta + \sin\theta\cos\theta\right)\tan\phi - \sin^2\phi}{2\left(\left(1 - \left(\frac{\pi}{2} - \phi - \theta + \tan\theta\right)\tan\phi\right)\right)\cos\theta} \tag{62}$$

## 5. Comparison of solutions with the previous studies

Under scaled stress analysis, similar patterns of stress distributions are assumed at all depths. This means the scaled stress variable $\chi$ and frictional variable $\beta$ are varied only with the slope of interest specified by the angular coordinate $\theta$. Due to the rigidly-perfect plastic assumption and scaled stress analysis, the only required mechanical parameter of material is the angle of friction $\phi$. The derived forms of exact solution for $\chi$ and $\beta$ are compared with Nadai's original/alternative solutions and Marais's corrected solution as shown in **Table 1**.

Unlike other solutions where $\beta=\sin\phi$, states of stress under Nadai's criterion do not reach a fully mobilized state of active condition everywhere. At the sliding plane where $\theta=\theta_f$, $\beta$ in Eq.(62) is undetermined because of the singularity, but by taking limit $\theta\rightarrow\theta_f$, it is found that $\beta$ approaches to $\sin\phi$. At the central plane where $\theta=\theta_c$, $\beta$ in Eq.(62) is less than $\sin\phi$. Therefore, an entire sand heap under Nadai's criterion stays in an admissible state just below yielding criteria because $\beta<\sin\phi$ in sand heap and $\beta\rightarrow\sin\phi$ at the slope surface only; i.e. the range of $\beta$ is varied between $\beta_c$ and $\beta_f$.

$$\beta_f = \lim_{\theta\rightarrow\pi/2-\phi}\beta = \sin\phi \tag{63}$$

$$\beta_c = \beta\big|_{\theta=0} = \frac{\cos^2\phi}{2\left(1-(\pi/2-\phi)\tan\phi\right)} - \frac{1}{2} \tag{64}$$

Stress solutions for a planar heap in a rectangular coordinate system are obtained by substituting for $\omega$ from Eq.(28) to Eqs.(16)-(18). Depth of interest $z$ and a physical parameter of unit weight $\gamma$ are likewise required.

$$\sigma_x = \gamma z\chi\frac{1-\beta\cos\theta}{\cos\theta} \tag{65}$$

$$\sigma_z = \gamma z\chi\frac{1+\beta\cos\theta}{\cos\theta} \tag{66}$$

$$\tau_{xz} = \gamma z \chi \beta \frac{\sin\theta}{\cos\theta} \tag{67}$$

The vertical thrust $P$ acting on the half-width base determined by integrating Eq.(66) over the base of the half-width wedge is calculated to check the weight-balanced condition under Nadai's criterion. By assigning the height of the ridge over the base $h$ to z with $\chi$ from Eq.(61) and $\beta$ from Eq.(62), the exact solution satisfies the weight-balanced condition by achieving $P=W$.

$$P = \int_0^{\pi/2-\phi} \sigma_z \frac{h}{\cos^2\theta} d\theta = \frac{1}{2}\gamma h^2 \cot\phi = W \tag{68}$$

Generally, a stress distribution with the property that the total normal pressure beneath the sand heap equals the weight of the heap is the immediate consequence of any law obeying continuum force balance under traction-free slopes. This fact indicates that some past approaches lack this property due to an ill-posed configuration of equilibrium equations and incorrect boundary conditions specified to the problem.

All solutions referring to $\phi$=30$^{\text{o}}$ are computed and plotted in **Figs.8-10**. **Figs.8** and **9** illustrate the distribution profiles of $\chi$ and $\beta$ along an arbitrary circular arc length $s$ normalized by its maximum at the sliding plane $s_f$. In spite of the insignificant differences, **Fig.8** reveals that the stress solutions at the central plane given by Eqs.(34) and (39) overvalue the exact solution, while that of Eq.(35) undervalues it. **Fig.9** reveals that the magnitude of $\beta$ is gradually decreased from the maximum value at the sliding plane and asymptotically approaches the minimum value at the central plane.

In **Fig.10**, Mohr stress circles were drawn to exhibit states of stress underneath the sand heap for one unit depth. Various positions were selected along the horizontal plane, ranging from the toe to the center line of the sand heap. The center and radius of each Mohr circle are calculated from $p$ and $q$ in Eqs.(13) and (14), respectively, using stress components in the rectangular coordinate obtained from Eqs.(65)-(67). The states of stress for the toe were represented by a Mohr circle having zero radius, while states of stress for the centerline were represented by a Mohr circle with the largest radius. All Mohr stress circles were confined within the failure envelope sloped by tan$\phi$. Consequently, the sand heap is stable and immobilized below the limit state.

## 6. Validation of solutions with the experiments

Stress distribution at the base of a granular heap as revealed by experimental studies has recently elicited much interest in the field of granular physics. Measuring pressure underneath a sand heap in both cone and wedge geometries as conducted by Hummel and Finnan (1921) appears to be the earliest experiment conducted, revealing that the basal pressure does not follow the sand heap's shape but exhibits a nonlinear distribution. Conical sand heaps formed by pouring sand at various heights spanning from 4.5 to 14.5 cm with 1 cm increments on each of the stages were conducted by Vanel et al. (1999). Their systematic measurements confirm that the pattern of stress distribution in the sand heap is convincingly independent of height. Therefore, stress profiles scaled by the product of unit weight $\gamma$ and maximum height of sand $h$ are typically plotted with the horizontal position from the ridge $x$ normalized by the length of the half-width base $h\cot\phi$.

Experimental evidences for conical, triangular and trapezoidal geometries of granular heaps constructed by various methods of sand deposition on rigid or deflected bases in dry condition can be found in a wide array of research works (Brockbank, et al., 1997; Geng, et al., 2001; Hummel and Finnan, 1921; Jotaki, 1979; Lee, 1956; Lee and Herington, 1971; Marais, 1969; Smid and Novosad, 1981; Trollope, 1956; Vanel, et al., 1999; Wiesner, 2000). Many experiments measured vertical normal pressure profiles along the base, but actual measurements of shear stresses are rare. None of the published experiments measures horizontal normal pressure along the central plane, due to limitations in experimental technique. Explanations on the three typical formation methods: layered sequences, wedge sequences and reposed sequences, are outlined in **Fig.11**.

According to the experimental results given by Lee and Herington (1971), it can be observed that the sand deposition method in layered and wedge sequences did not significantly affect the stress profiles measured at the final stage. This outcome provides good support to the assumption of below-limiting stress used in this study because it can point out that the formation of a sand heap is an elastic stage independent of formation history. However, the effect of formation would appear, and the theory applied in this study would prove inadequate if the sand heap is constructed using reposed sequences as employed by Vanel et al. (1999) thru the localized source procedure. Under reposed sequences, a sand heap grows in size without any change in proportional shape throughout the formation process because sand deposited in thin layers on the sliding face always cascades down to build a new slope, which remains inclined at the angle of repose. Therefore, a whole sand heap buried consecutively might undergo the plastic stage with the fixed principal axes (Wittmer, et al., 1997; Wittmer, et al., 1996). With this simple conclusion that is different from Nadai's criterion, solving the static equilibrium conditions result in a phenomenon of central stress dip (Watson, 1991; 1996) due to the arch action, as explained by Michalowski and Park (2004).

We shall restrict ourselves to comparing the exact solutions with the available experimental data of long triangular embankments loosely placed on a rigid base (wedge sequences with sand $\phi$=32.5$^{o}$ as per Lee (1956), layered and wedge sequences to $h$=38.1 cm with sand $\phi$=30$^{o}$, $\gamma$=15.02 kN/m$^{3}$ as per Lee and Herington (1971), wedge sequences to $h$=38.1 cm with sand $\phi$=30$^{o}$, $\gamma$=14.9 kN/m$^{3}$ as per Wiesner (2000), reposed sequence to $h$=43.18 cm with sand $\phi$=32.5$^{o}$, $\gamma$=15.20 kN/m$^{3}$ as per Hummel and Finnan (1921), layered and reposed sequences to $h$=8 cm with sand $\phi$=33$^{o}$ as per Vanel et al. (1999), wedge sequences to with sand $\phi$=32.5$^{o}$ and crushed aggregate $\phi$=40$^{o}$ as per Trollope (1956) and Trollope and Burman (1980)). Herein, a local error on the measurement of middle strips was reported due to a slight distortion of the supporting metal plate in Lee and Herington (1971).

In **Fig.12**, stress profiles of scaling vertical stress $\sigma_z/\gamma h$ and scaling horizontal shear stress $\tau_{xz}/\gamma h$ plotted against a normalized half-width base $x/(b/2)$ exhibited an acceptable agreement with the exact solution, taking $z=h$ and $x/(b/2)=\tan\theta/\tan\theta_f=\tan\theta/\cot\phi$. The exact solution presents a hump profile of vertical pressure with no central stress dip. This result is consistent with Savage's (1998) argument against the closure regarding fixed principal axes, that there is no clear stress dip in prismatic heaps; therefore, the maximum vertical pressure $\sigma_{z,\max}$ can be found at the center.

$$\sigma_{z,\max} = \sigma_z\big|_{\theta=0} = \gamma h \frac{1+\cos^2\phi - (\pi/2-\phi)\tan\phi}{2\cos^2\phi} \tag{69}$$

Although the match between the analytical solution and the past experimental data is reasonably good, the generality of this result might not be readily persuasive as the limited extent of $\phi$ is shown.

Moreover, the horizontal stresses measured beneath the model embankment are not available. According to **Fig.12**, the distributions of vertical stress and shear stress obtained from both theory and experiment, for $\phi$ between $30^{o}$-$40^{o}$ lie chiefly within a narrow range of variation; however, the distribution of horizontal stress obviously stretches out in a wider range. As a result, the horizontal stress distribution is a severe test of the effectiveness of the theory; therefore, a more stringent experiment is required to validate whether the predicted stress distribution on the base varies with $\phi$ in a manner consistent with experiments.

Since the theoretical solution stems from a continuum theory of stresses throughout the heap, validation of the theory by demonstrating that predicted stresses in the bulk are also in good agreement with the experiments would be an important aspect of making the result more convincing. Nevertheless, it is not simple to measure stresses in the bulk experimentally because the pressure cells/strain gauges are basically attached to the base before the stage of sand deposition.

## 7. Remarks on closures of stress relation

The relation described in Eq.(20) can be rearranged with the angle of principal compressive stress $\omega$ as an unknown variable. If the relation of $\omega$ is given, this stress relation appears to link all stress components that might give rise to a misinterpretation of a constitutive equation.

$$\sigma_x = \sigma_z - 2\left|\tau_{xz}\right| / \tan 2\omega \qquad (70)$$

Generally, constitutive equations of materials are expressed in terms of stress invariants and material parameters, and they are independent of boundary conditions. Therefore, Eq.(70) is not a constitutive equation but is, rather, a closure or conjecture of the stress relation particularly applied to the specific problem, and not a universal problem. In this context, Savage (1997; 1998) explained that several papers relevant to stress distribution in sand heap actually proposed closures instead of constitutive equations. Consequently, Nadai's criterion is also regarded as a specific closure because the particular condition of a sand heap described by Eq.(28) is embedded in Eq.(70).

In contrast with the criterion of incipient failure everywhere (Sokolovskii, 1965; Wittmer, et al., 1997) where a whole sand heap is in a yielding state, the solution of stress obtained under Nadai's criterion is classified as statically admissible stress states below the yielding limit. Numerous stress solutions also fall under this class (Michalowski and Park, 2004) including the closure of fixed principal axes (FPA) proposed by Wittmer et al. (Wittmer, et al., 1997; 1996) as well as the general closure of oriented stress linearity (OSL) proposed by Wittmer et al. (1997).

Based on the FPA hypothesis, $\omega=\omega_f$ as defined by Eq.(27) is kept constant throughout a symmetric half of the sand heap and mirrored along the center plane to the other half of the heap (see **Fig.13**). Substitution of $\omega_f=\pi/4-\phi/2$ to Eq.(70) yields the closure of FPA in Eq.(71).

$$\sigma_x = \sigma_z - 2\left|\tau_{xz}\right|\tan\phi \qquad (71)$$

We will explain that Nadai's criterion based on Eq.(28) also provides the description of closure relation among stresses in extension to the scope of linearity limited by FPA to non-linearity. Nadai observed that visualization of his hypothesis on associating the major principal axis to the angular coordinate can be physically explained using circles. The orientation of the major principal stress along any circle of constant radius traced around the ridge as the center simply converges toward the same point situated at the top of the circle. In other words, the rays of major principal axes along the same circle intersect one another at the pole of the circle, so these major principal axes are polarized. (see **Fig.14**)

One might find that unlike the criterion governed by FPA where $\omega=\pi/4-\phi/2$ is fixed everywhere, the different concept in Nadai is where $\omega=\theta/2$, varying with the angle $\theta$ traced about the ridge. It is possible that his closure has not been noticed in many researches devoted to the subject in the late 1990s, with a focus on the phenomenon of arching in a sand pile (Cantelaube and Goddard, 1997; Cates, et al., 1998; Didwania, et al., 2000; Wittmer, et al., 1997; Wittmer, et al., 1996). Nadai's criterion gives rise to a new closure relation coined to polarized principal axes (PPA), owing to its characteristics that match with the existing closure of fixed principal axes (FPA). A clear difference in the criteria of principal axes under FPA and PPA can be observed by comparing **Fig.13** with **Fig.14**.

Substitution of $\omega=\theta/2$ under Nadai's criterion to Eq.(70) obtains the closure of PPA in Eq.(72). This equation of stress relation is not characterized by stress invariants and material parameters. Local coordinates are also embedded, so the stress relation under PPA is definitely not a constitutive equation. Rather, it is a closure that is entirely based on hypotheses with no concrete mechanical and experimental evidences, similar to FPA. Eq.(72) is comparable with Eq.(71), so the closure of PPA can be reduced to the closure of FPA by specifying a term $z/x$ to $\tan\phi$ representing the proportional shape of the sand heap.

$$\sigma_x = \sigma_z - 2\left|\tau_{xz}\right| z/x \qquad (72)$$

## 8. Discussions on stress distributions

A fundamental equation derived in Appendix A provides the second-order differential equation in general form, which can be applied to the statics of a planar sand heap, not only for PPA but also FPA and other related closures. Substitution of $\omega=\pi/4-\phi/2$, described by FPA, to Eq.(8) gives $\psi=\pi/4-\phi/2-\theta$; therefore, $\psi'=-1$ and $\psi''=0$. According to Eqs.(A.9)-(A.11) shown in Appendix A, coefficients $A=0$, $B=1$ and $C=0$ are obtained, reducing Eq.(A.8) to the fundamental equation for FPA which is ready to be solved using the boundary conditions in much the same way as the solution under PPA.

$$\chi'' + \chi = 0 \qquad (73)$$

The stress distributions under FPA for a planar heap are, in bi-linear shapes, characterized by an unrealistic drop in vertical stress at the center line (Wittmer, et al., 1997; Wittmer, et al., 1996). Still, there is an abundance of closures that can give a statically admissible stress solution by satisfying equilibrium conditions, below-limit stress criteria and boundary conditions of sand heap. Closures for OSL (Wittmer, et al., 1997) and illustrative models given by Michalowski and Park (2004) are examples. We realize that the statics of sand heap take the form of many solutions with/without an indication of formation dependence. So this problem is not as simple as the geometry of sand heap itself, which has a unique solution of shape sloped by the angle of repose.

Still, the closure of PPA has a limitation to describe stress distribution in a sand heap where the deposition history is unclear or might be more complicated than the typical formations. Though the effects of preparation history on the resultant stress distribution was discussed earlier to clarify the types of preparation which the PPA closure is expected to work for, it is unlikely these conditions are intrinsically present in an arbitrary sand heap. Since the present analysis is primarily based on continuum mechanics of uniform materials, the analytical stresses would differ to the actual stresses because the effects of grain size distribution, non-uniform friction angle, grain shape and angularity, variation of bulk density and properties of individual grains are not considered.

Typical features of stress distributions characterized by PPA are shown in **Figs.15-16** using $\phi$=30°. The nonlinear pattern of weight transmission can be observed by the contour of $\sigma_z/\gamma h$ in **Fig.15**. Vertical pressure is intense around the center of the sand heap and is gradually lower toward the toes. Most of its own weight is transferred to the central portion, where a margin of sliding resistance is greater than that of the toe, which almost slides at the sliding surface. The variations in principal stress orientations shown in **Fig.16** can indicate that compression, loaded by its own weight, is carried along principal compressive directions. These trajectories can be regarded as stacks of curved arches, not necessarily straight arches as characterized by FPA or OSL in Witter, et al. (1997).

## 9. Summary

The findings in this study can be summarized as follows:

(1) Classical theories contributed by Nadai (1963) provide sources of background in a wide range of mechanics. Nadai's sand hill analogy emphasizes his achievements in both structural mechanics and geo-mechanics. Although, there is no physical implication to associate the ultimate torque of a twisted member to stress distributions beneath a sand heap growing on the cross-section, some facts emanating from Nadai's sand hill analogy elicit considerable attention particularly regarding geometry and the statics of sand heaps. On the aspect of geometry, it was found that the maximum volume of sand that can be loosely deposited on a given cross-section of the rigid base is almost independent of the methods and sequences of heap formation because the final shape and stable height are regulated by the angle of repose. However, this is not true when it comes to statics because stress distributions underneath the sand heap indicate strong dependence on formation history. Therefore, the solution for geometry is unique but is varied for statics. At this point, we found that there is no problem in using Nadai's sand hill analogy in the calculation of ultimate torque from the maximum volume of sand poured on that cross-section.

(2) Actually, the aspects of geometry and statics are mutually interactive; i.e. a geometry of sand

heap provides a boundary condition to the static problem while statics provide stability of shape to the geometrical condition. In light of the natural law of friction, Nadai's sand hill analogy is accompanied by Nadai's sand heap criterion with regard to the stability of sand heaps in order to support the existence of the unique sand heap shape. Therefore, theoretical stress distribution based on Nadai's criterion could be derived under static equilibrium and boundary conditions of stable sand heap shapes. Nadai failed to achieve this validation because the derived stress solution violates the equilibrium conditions. We found that the source of error is not Nadai's criterion. It was rooted in the method of derivation, which was still limited to a certain concept. This concept was ruled out by recent developments in plasticity theory.

(3) The statics of a vertically symmetrical loose triangular embankment having sliding planes inclined at the angle of repose was introduced in this study as a demonstrative problem to be validated. The present paper is primarily concerned with Nadai's sand heap criterion, as applied to the problem of initial stresses in a planar sand heap. Its review of the theoretical background elaborates the error found in this classical theory and delivers a correction based on the theory of continuum mechanics. We found that the error stems from an ill-posed problem in the configuration of equilibrium equations. The number of equations is greater than the number of variables by one. Because both Nadai (1963) and Marais (1969) drop one equation in order to achieve balance with the number of variables, their solutions clearly violate one of the equilibrium conditions.

(4) In the recent work of Michalowski and Park (2004), the broader framework of plasticity theory, which can handle below-limiting stress was employed to formulate statically admissible stress fields that do not violate the failure condition and that satisfy both equilibrium requirements and stress boundary conditions. We fixed the equation systems by introducing a frictional variable that is dependent on location to replace a constant frictional parameter. This frictional variable does not represent the variable property of materials but rather, a gradient function describing an admissible state of stresses. This technique appeared to be an effective method of solving the problem under Nadai's criterion.

(5) By following regular procedures employed by Sokolovskii (1965), the system of equations were manipulated in scaled stress. A fundamental equation in the form of second-order ordinary differential equation was formulated. This equation has one order higher than the fundamental equation formulated in earlier studies. According to the characteristics of this differential equation, we found that the exact closed form can be derived. Two stress boundary conditions were necessary. We took one primary condition from zero-traction along the sliding surface. We also suggested taking a supplementary condition in terms of ordinary derivatives along the same sliding surface.

(6) The admissible stress solution was verified through the equilibrium conditions and weight-balanced condition. The vertical stress distribution along the horizontal base indicated a hump profile with the maximum pressure at the heap's center. States of stress along the horizontal base represented by Mohr circles confirmed that stresses in the sand heap did not violate the failure criterion and stayed below the limit equilibrium.

(7) Validations with the published experiments on sand undertaken during the years 1921–2000 by various researchers were made. In addition to vertical normal stresses, horizontal shear stresses along the base were measured by Lee and Herington (1971) as well as Wiesner (2000). Comparison of the exact solution and the measured value yielded acceptable agreement. Moreover, we found that the approach of Nadai's criterion might be suitable for static sand heaps constructed from a spread source rather than a line source because sand would not be in a plastic state, such that the influence of complicated formation history on stress distribution is insignificant.

(8) Because the orientation of the major compression stipulated in Nadai's criterion could be regarded as the new closure of stress relation, the closure of polarized principal axes (PPA) was termed to match its characteristics with the existing closure of fixed principal axes (FPA) proposed by Wittmer et al. (Wittmer, et al., 1997; 1996). We might infer that self-weight transmission under Nadai's closure is in a state of equilibrium of the compressive forces holding the configuration of the arches together in both major and minor directions below limiting stress. These stacks of arches assembled into columns of curves dominantly transfer outward lateral pressure and its own weight downward to a rigid base supporting the sand heap.

## 10. Conclusion

In conclusion, the analyses and implications drawn from this study are related to Nadai's hypotheses from his classical text books. The theoretical formulas for the exact distribution of stresses based on Nadai's criterion were rigorously derived in accordance with the continuum theory, and findings sufficiently agreed with the experimental measurements. The results of the present study permitted the acceptance of Nadai's sand hill analogy in relation to the stable shape and statics of a sand heap sloped at the angle of repose. The polarized principal axes hypothesis that originated from Nadai's criterion presents a new closure scheme for the sand heap. Though Nadai's closure is not physically motivated and its approach does not shed any insight onto the granular physics involved, it can provide a convenient empirical assertion in providing the simplified stress solutions and initial stresses to various kinds of engineering works including earth dams, embankments, stockpiles and waste dumping because triangular-shaped sand heaps under self-weight loading is a fundamental problem in geo-mechanics. More research efforts into this subject can be extended to the trapezoidal shape, conical shape and other different shapes of sand heaps, as well as weight-transferred mechanisms and arch actions in granular media.

## Acknowledgements

This research was conducted under the Grants-in-Aid for Scientific Research No.21360225 funded by the Japan Society for the Promotion of Science. The authors would like to express their appreciation to the society for supporting this fundamental study.

## Appendix

### A General formulation of fundamental equation

The second derivatives $\chi''$ based on Eq.(42) is obtained by chain rule differentiation considering $\chi'=\chi'(\theta,\chi(\theta),\beta(\theta),\psi(\theta),\psi'(\theta))$:

$$\chi''=\frac{d\chi'}{d\theta}=\frac{\partial\chi'}{\partial\theta}+\frac{\partial\chi'}{\partial\chi}\chi'+\frac{\partial\chi'}{\partial\beta}\beta'+\frac{\partial\chi'}{\partial\psi}\psi'+\frac{\partial\chi'}{\partial\psi'}\psi'' \tag{A.1}$$

where

$$\frac{\partial\chi'}{\partial\theta}=-\frac{\sin(2\psi+\theta)}{\sin 2\psi} \tag{A.2}$$

$$\frac{\partial \chi'}{\partial \chi} = -\frac{\cos 2\psi + (3+2\psi')\beta}{\sin 2\psi} \tag{A.3}$$

$$\frac{\partial \chi'}{\partial \beta} = -\frac{3+2\psi'}{\sin 2\psi}\chi \tag{A.4}$$

$$\frac{\partial \chi'}{\partial \psi} = 2\frac{\left(1+(3+2\psi')\beta\cos 2\psi\right)\chi - \cos\theta}{\sin^2 2\psi} \tag{A.5}$$

$$\frac{\partial \chi'}{\partial \psi'} = -\frac{2\beta}{\sin 2\psi}\chi \tag{A.6}$$

Because Eqs.(A.3), (A.5) and (A.6) are coupled with $\beta$, the expression $\beta$ can be rearranged from Eq.(42):

$$\beta = \frac{\cos(2\psi+\theta) - \chi\cos 2\psi - \chi'\sin 2\psi}{(3+2\psi')\chi} \tag{A.7}$$

Equation (A.1) substituted by Eqs.(A.2)-(A.6), using Eq.(43) for $\beta'$ and Eq.(A.7) for $\beta$, can be rearranged to the standard form of the second-order ordinary differential equation:

$$\chi'' + A\chi' + B\chi + C = 0 \tag{A.8}$$

where $A$, $B$ and $C$ are variable coefficients as summarized below:

$$A = 4(1+\psi')\cot 2\psi - \frac{2\psi''}{3+2\psi'} \tag{A.9}$$

$$B = -\left(3 + 4\psi' + \frac{2\psi''}{3+2\psi'}\cot 2\psi\right) \tag{A.10}$$

$$C = \frac{\cos(2\psi+\theta)}{\sin 2\psi}\left(4(1+\psi')\tan(2\psi+\theta) + \frac{2\psi''}{3+2\psi'}\right) \tag{A.11}$$

Invoking Nadai's criterion from Eq.(29) where $\psi=-\theta/2$, $\psi'=-1/2$ and $\psi''=0$ to Eqs.(A.9)–(A.11), $A=-2\cot\theta$, $B=-1$ and $C=0$ are obtained, thus reducing Eq.(A.8) to Eq.(49).

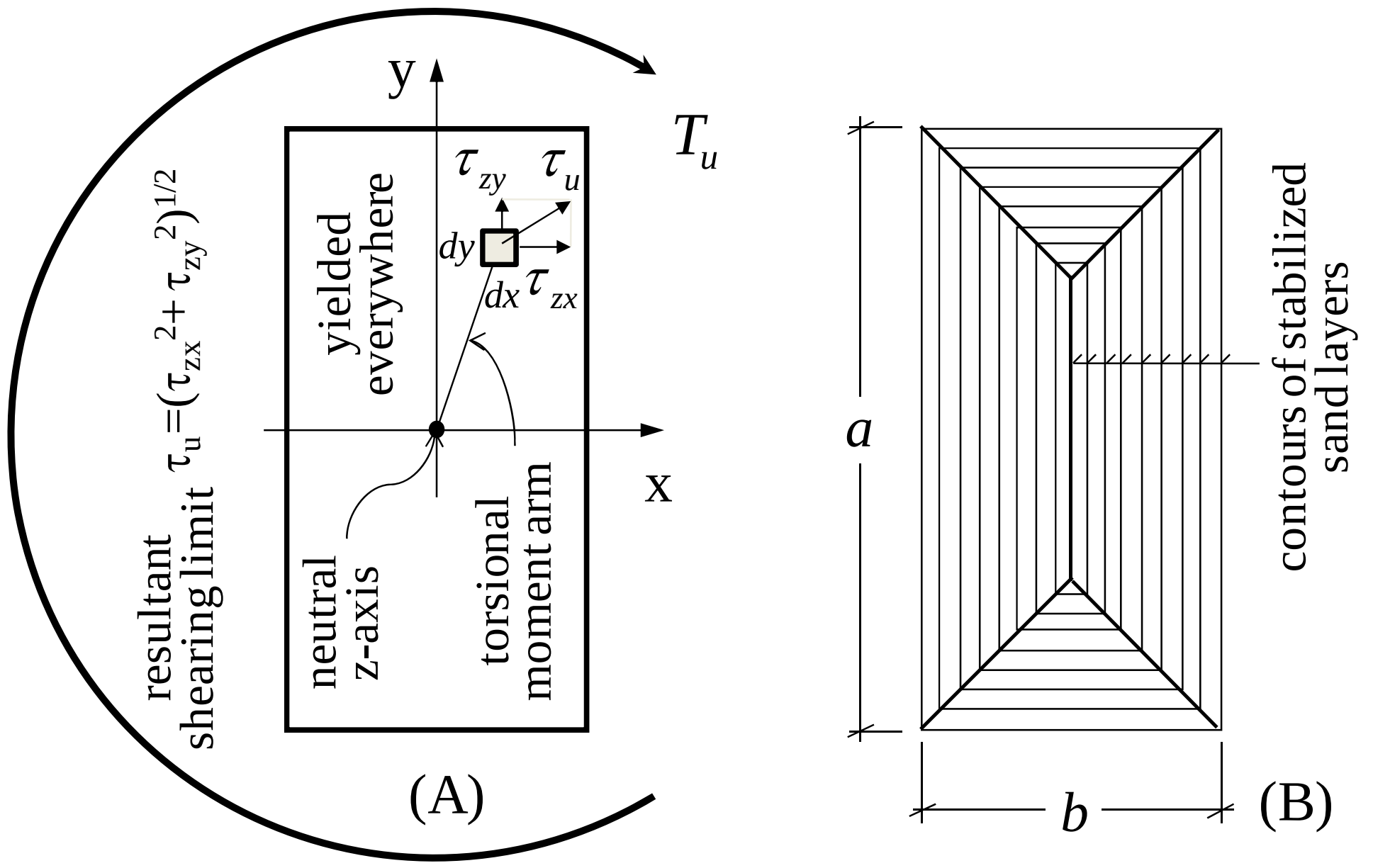

ultimate torsional moment $T_u = \int_x \int_y \left( x\tau_{zy} - y\tau_{zx} \right) dxdy = \frac{\tau_u}{6} b^2 \left( 3a - b \right)$

Figure 1 (A) State of shear stress under fully plastic condition due to ultimate torsional moment and (B) topography of sand heap piled into a rectangular cross-section.

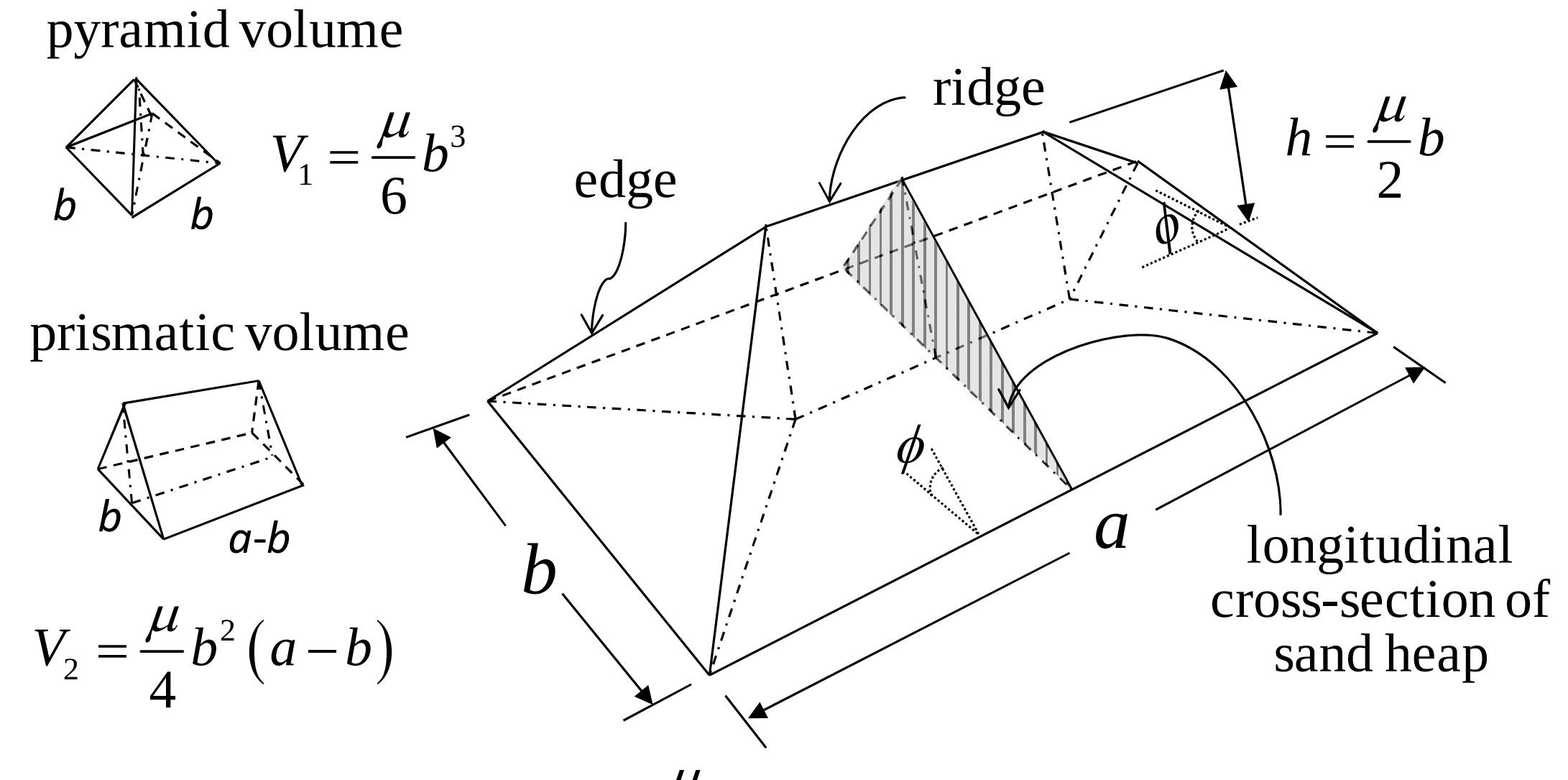

sand volume $V = V_1 + V_2 = \frac{\mu}{12} b^2 \left( 3a - b \right)$ where $\mu = \tan\phi$

Figure 2 Volume of sand heap deposited on a rectangular stiff base bounded by four stable slope surfaces and inclined horizontally along the angle of repose.

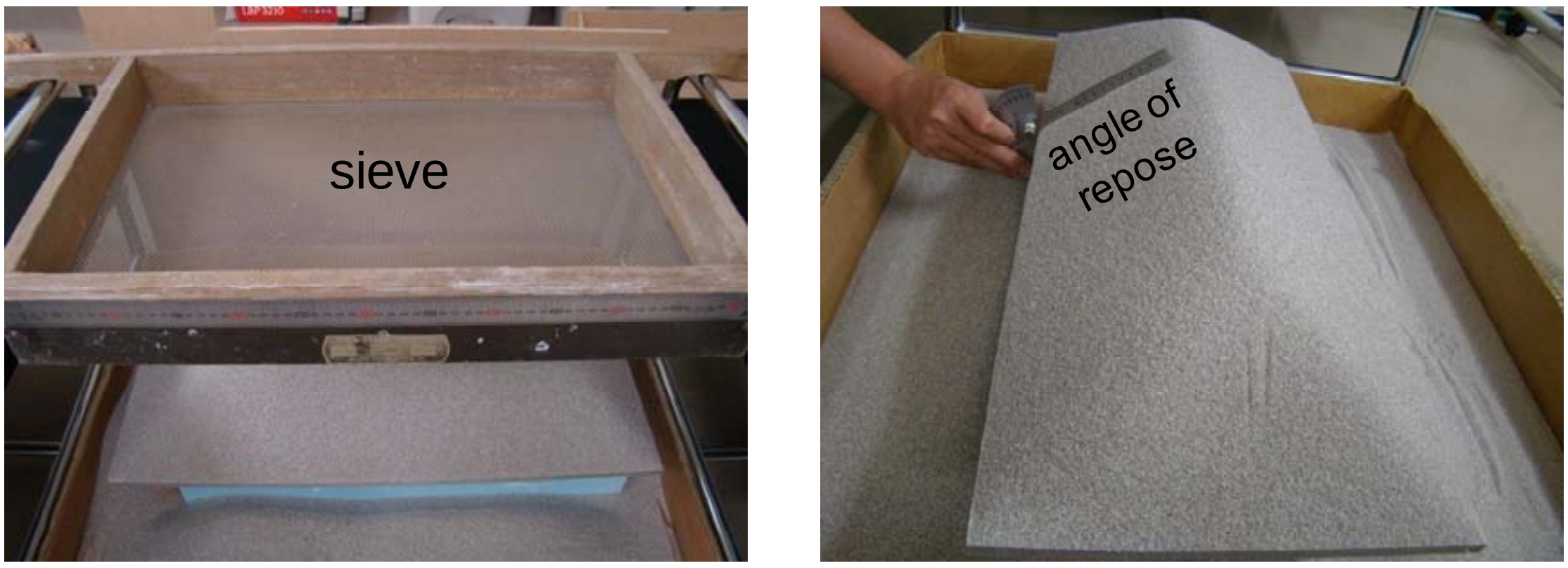

Figure 3 (Left) Sand heap constructed by pouring dry sand from a sieve source onto a stiff and rough rectangular base and (right) a final shape of the sand heap, with the angle of repose consistently observed along all sliding surfaces.

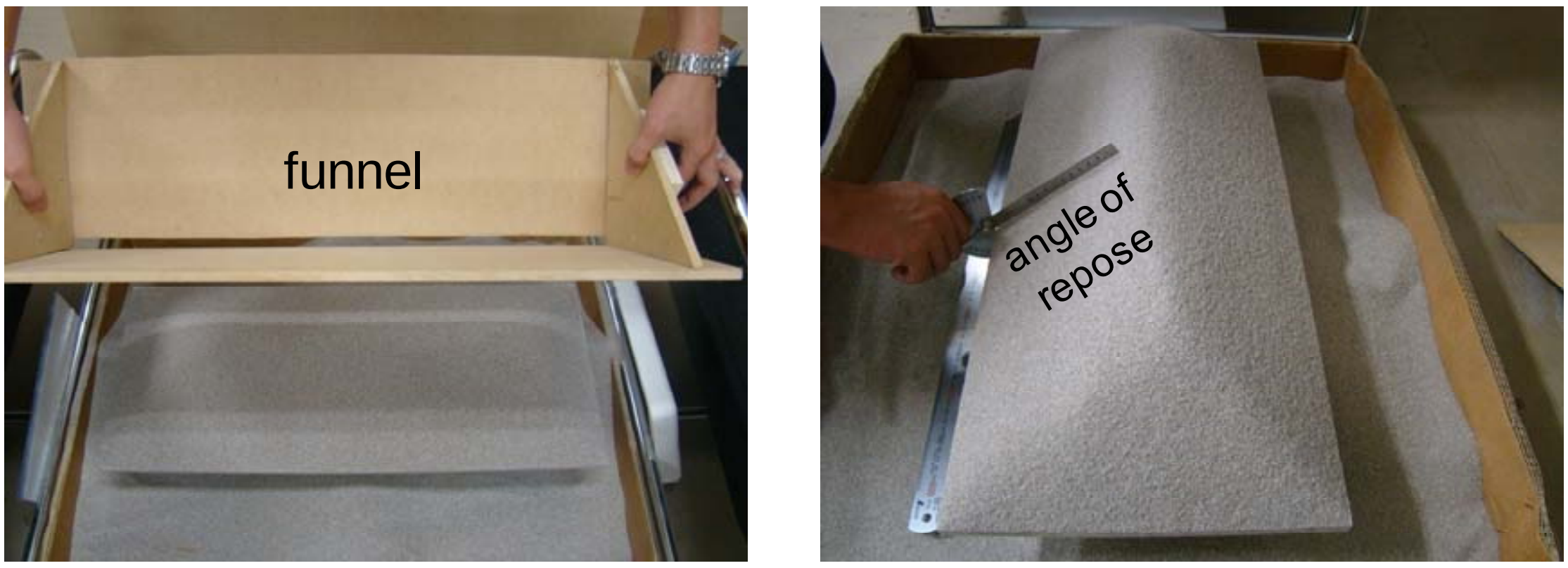

Figure 4 (Left) Sand heap constructed by pouring dry sand from a line source to a stiff and rough rectangular base and (right) a final shape of sand heap representing the angle of repose consistently observed along all sliding surfaces.

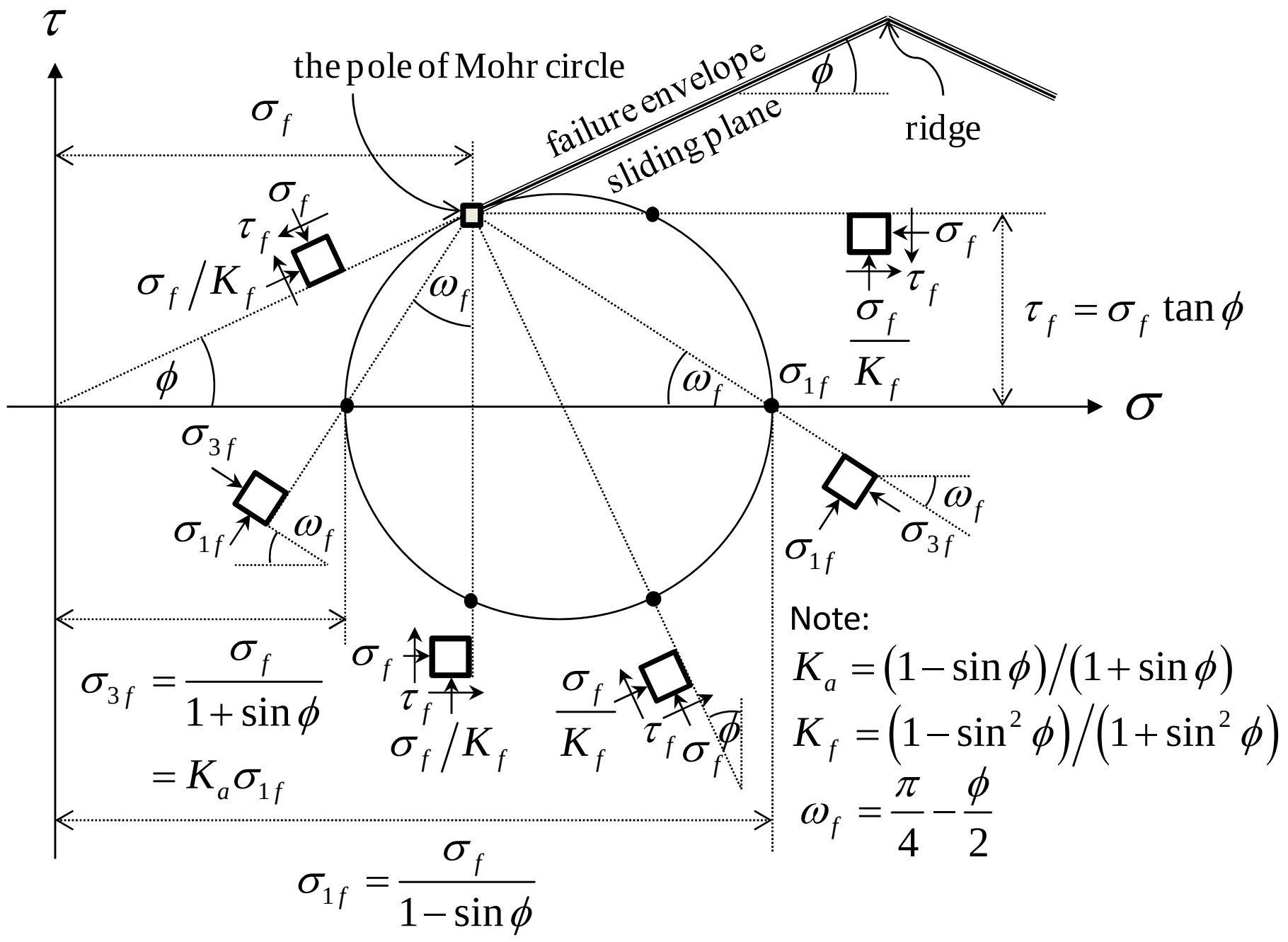

Figure 5 Typical Mohr stress circle describing the mobilized states of stress in various planes along the sliding plane in which inclination is taken to be parallel to the slope of the failure envelope.

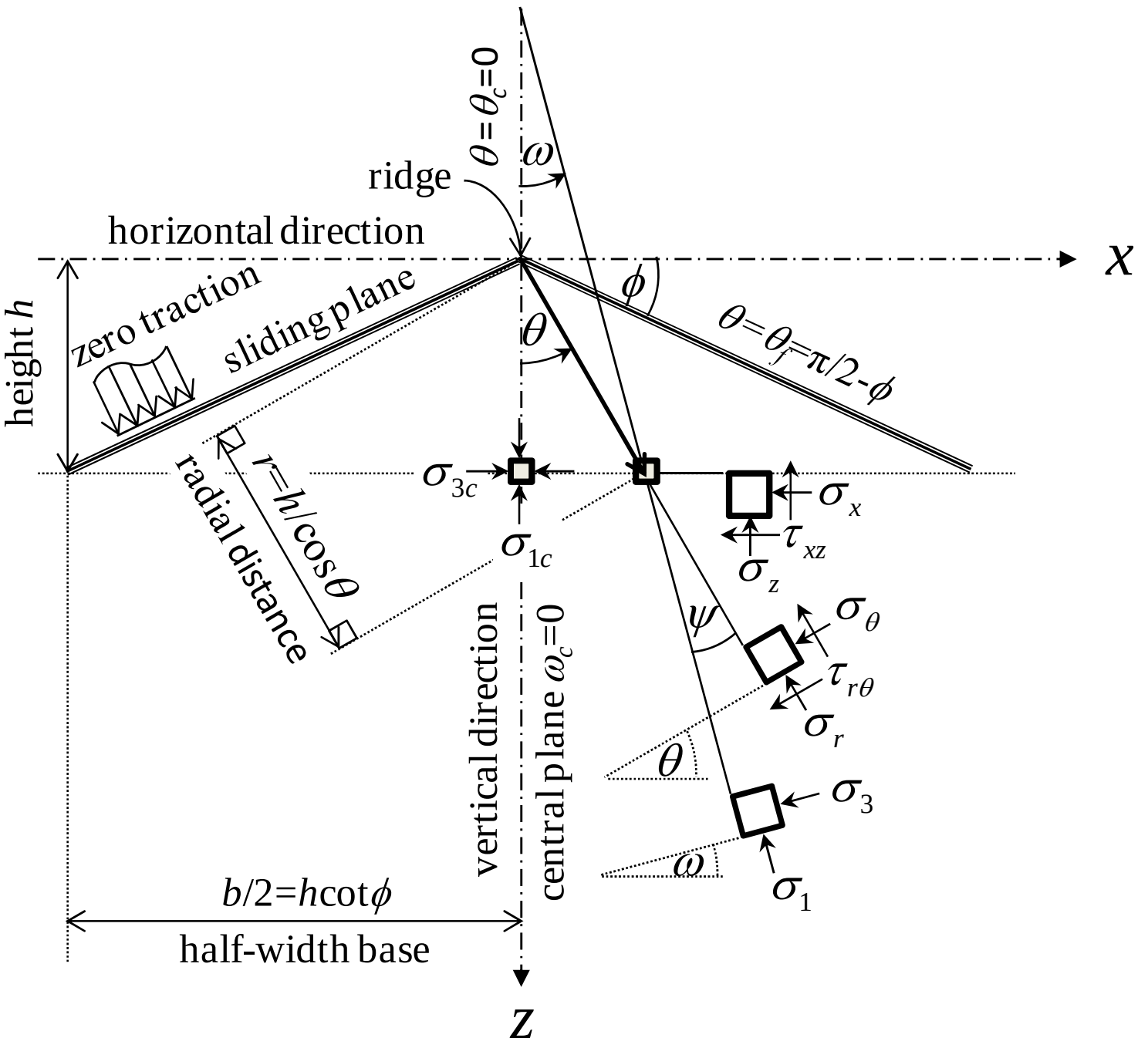

Figure 6 Geometry and boundary conditions of semi-infinite planar sand heap inclined at angle of repose with the horizontal where the referenced coordinate systems and states of stress described along various planes in a half-width symmetric side.

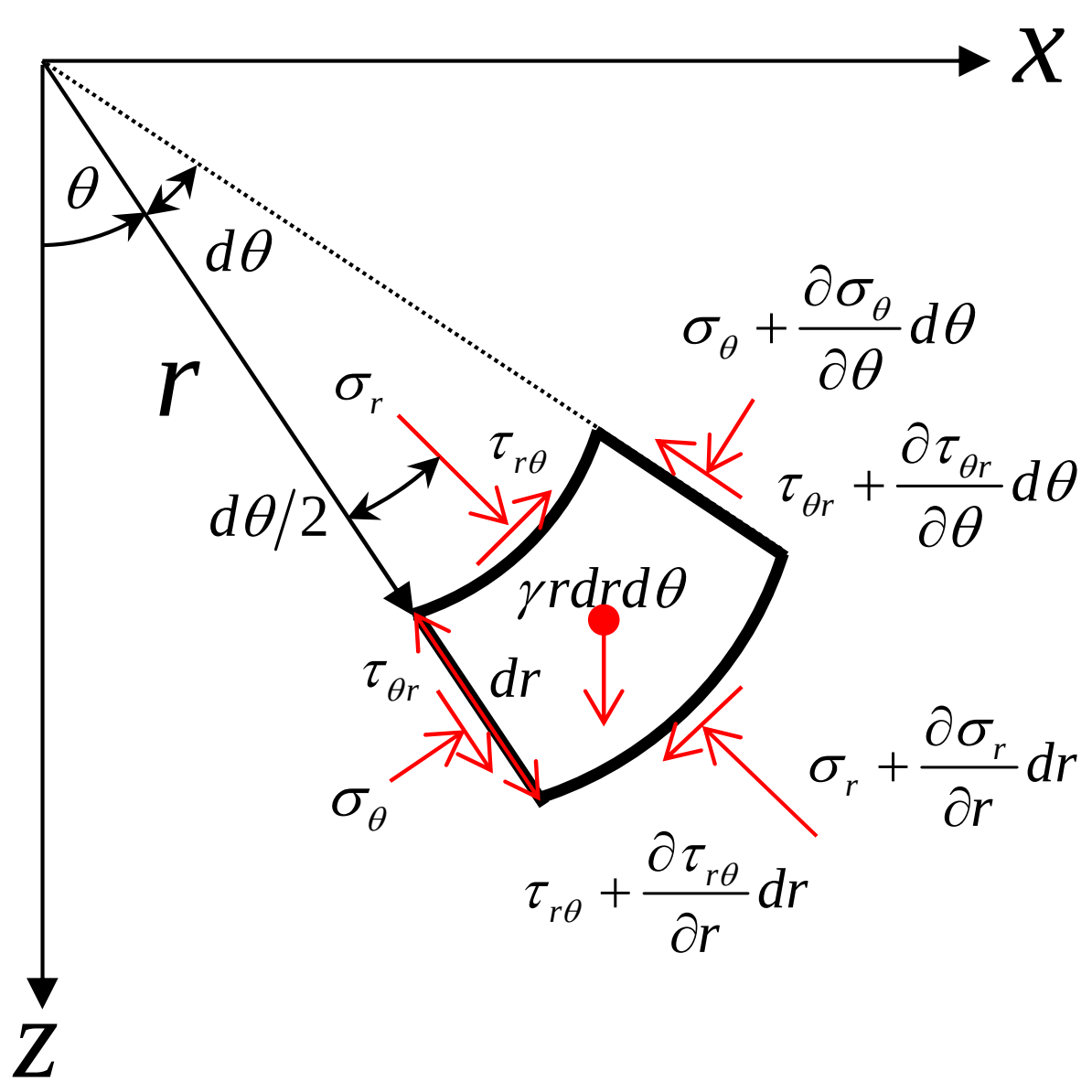

Figure 7 State of stresses on an infinitesimal body in polar coordinate ($r$,$\theta$) overlaid with rectangular coordinate ($x$,$z$).

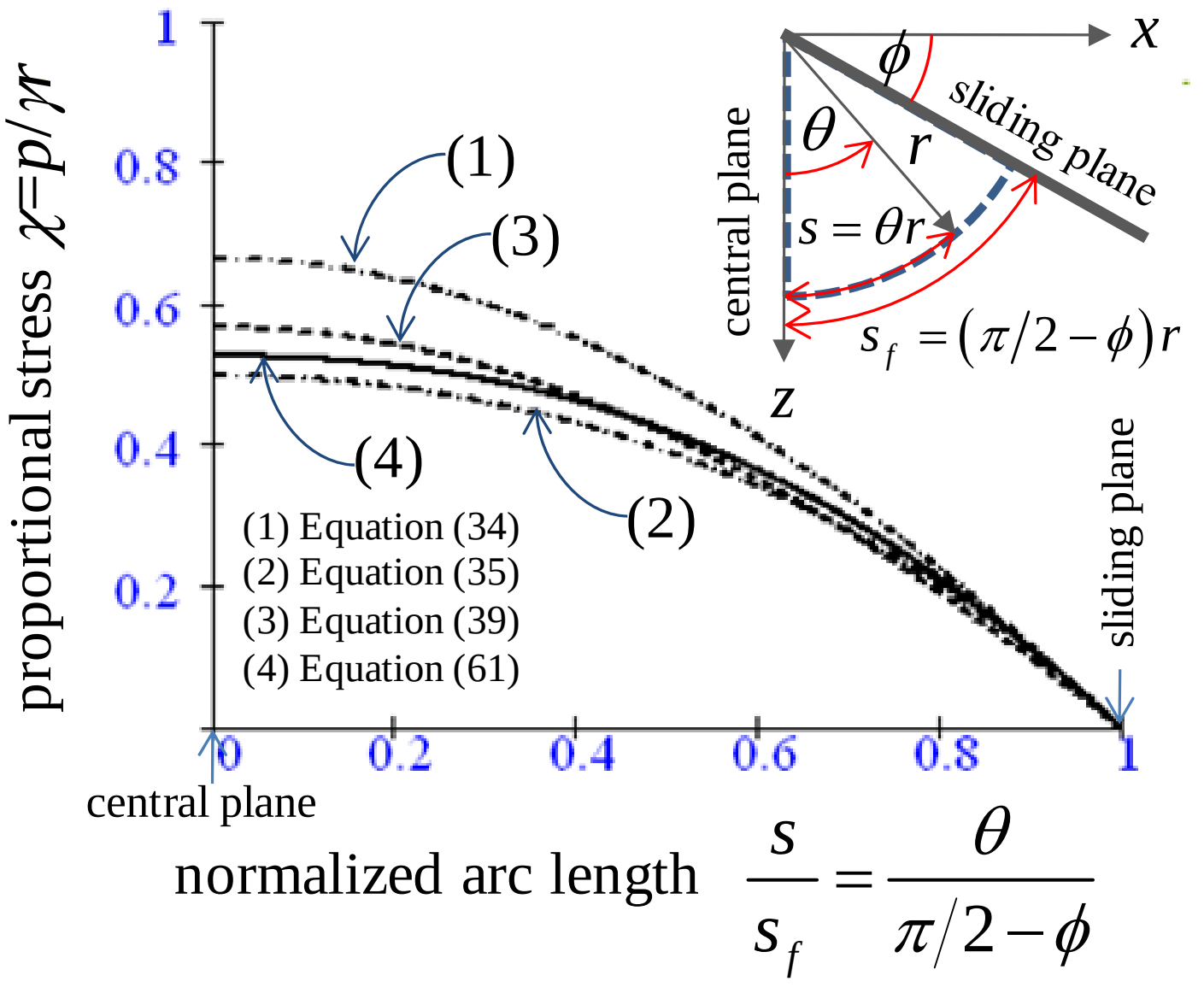

Figure 8 Comparison of various solutions for proportional stress $\chi$ with angular coordinate $\theta$ using $\phi$=30$^{\circ}$. (1) represents Nadai's original solution (1963), (2) represents Nadai's alternative solution, (3) represents Marais's corrected solution (1969) and (4) represents the exact solution (this study).

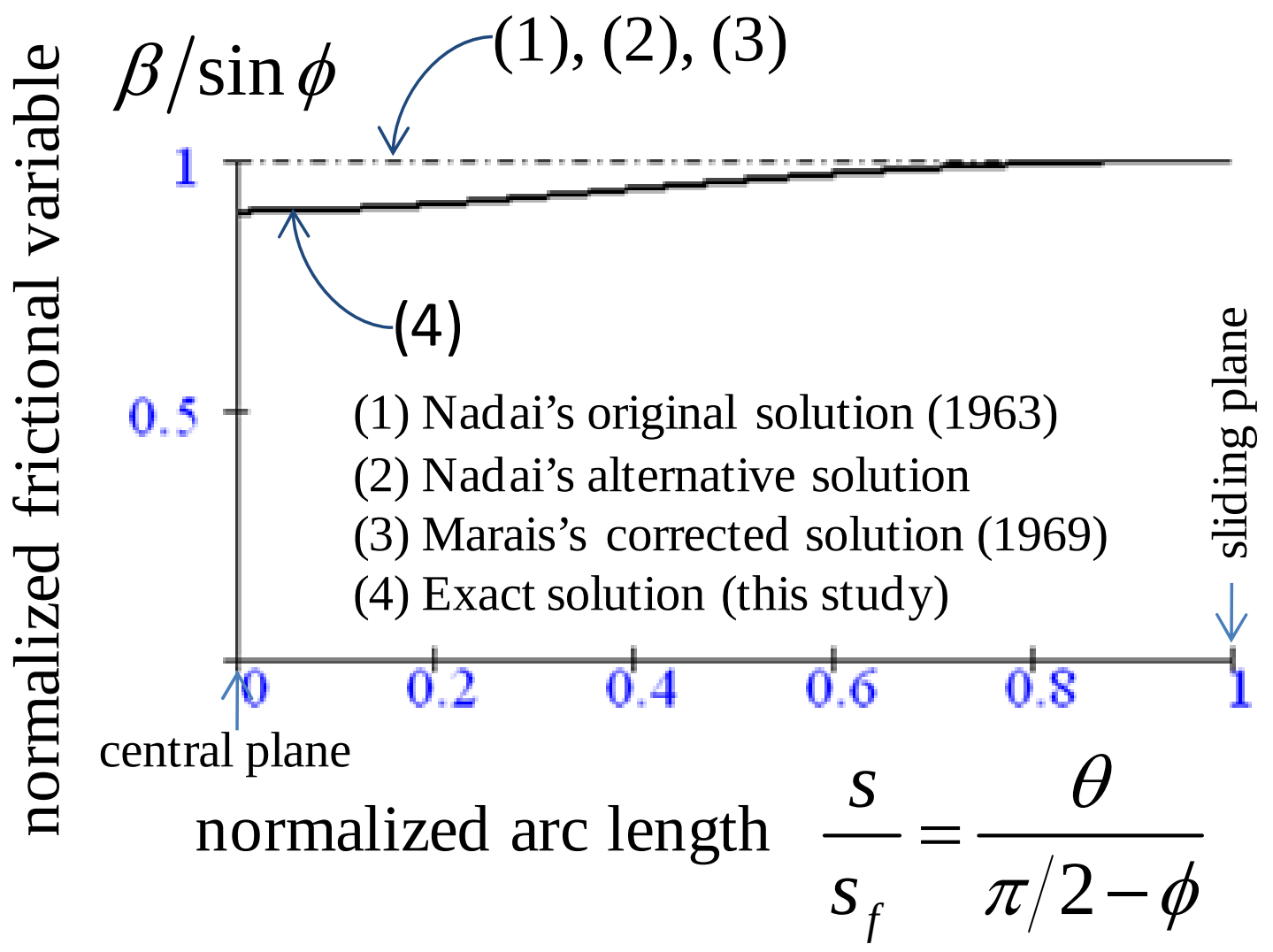

Figure 9 Comparison of various solutions for frictional variable $\beta$ with angular coordinate $\theta$ using $\phi$=30$^{o}$. Constant $\beta$=sin$\phi$ was used in the past studies. Nadai's closure relation is satisfied by letting $\beta$<sin$\phi$ vary as a function of angle in this study.

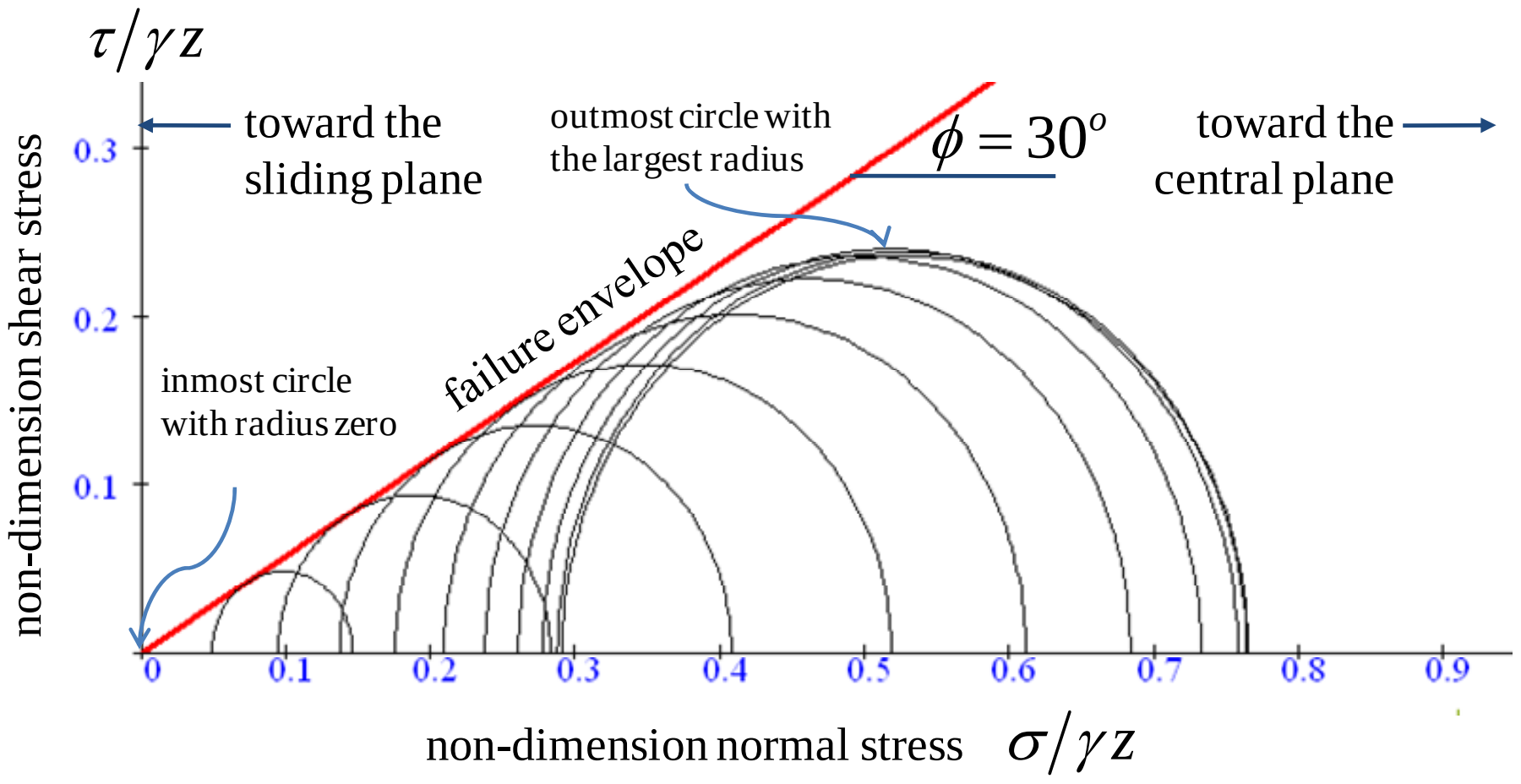

Figure 10 Mohr circles in non-dimension stress for a sand heap with $\phi$=30$^{o}$. Eleven circles represent state of stresses in the bulk at the same vertical height with equal horizontal interval away from the centerline. The smaller circles correspond to stresses located toward the sliding plane while larger circles correspond to stresses located toward the central plane. The circle with radius zero touching the failure envelope represents stress at the edge, and the biggest circle lying inside the failure envelope represents stress at the center.

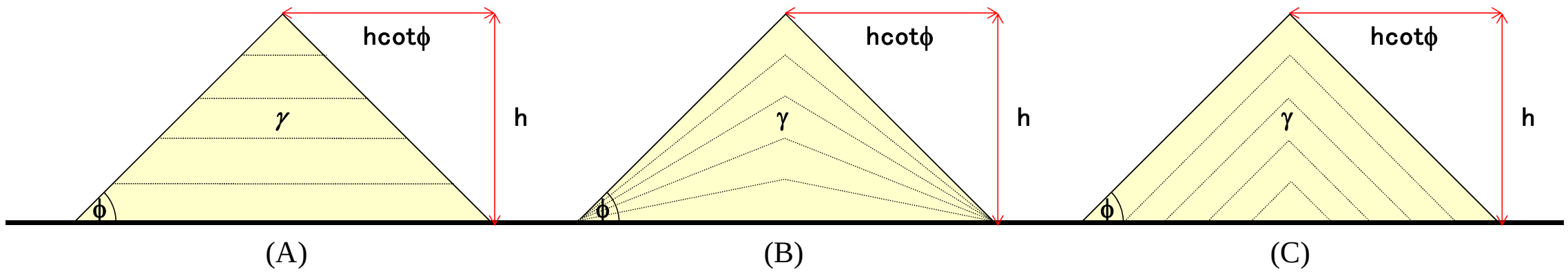

Figure 11 (A) Layered sequences built by the sieve method, (B) wedge sequences built by the scoop method and (C) reposed sequences in proportional shape built by the funnel method.

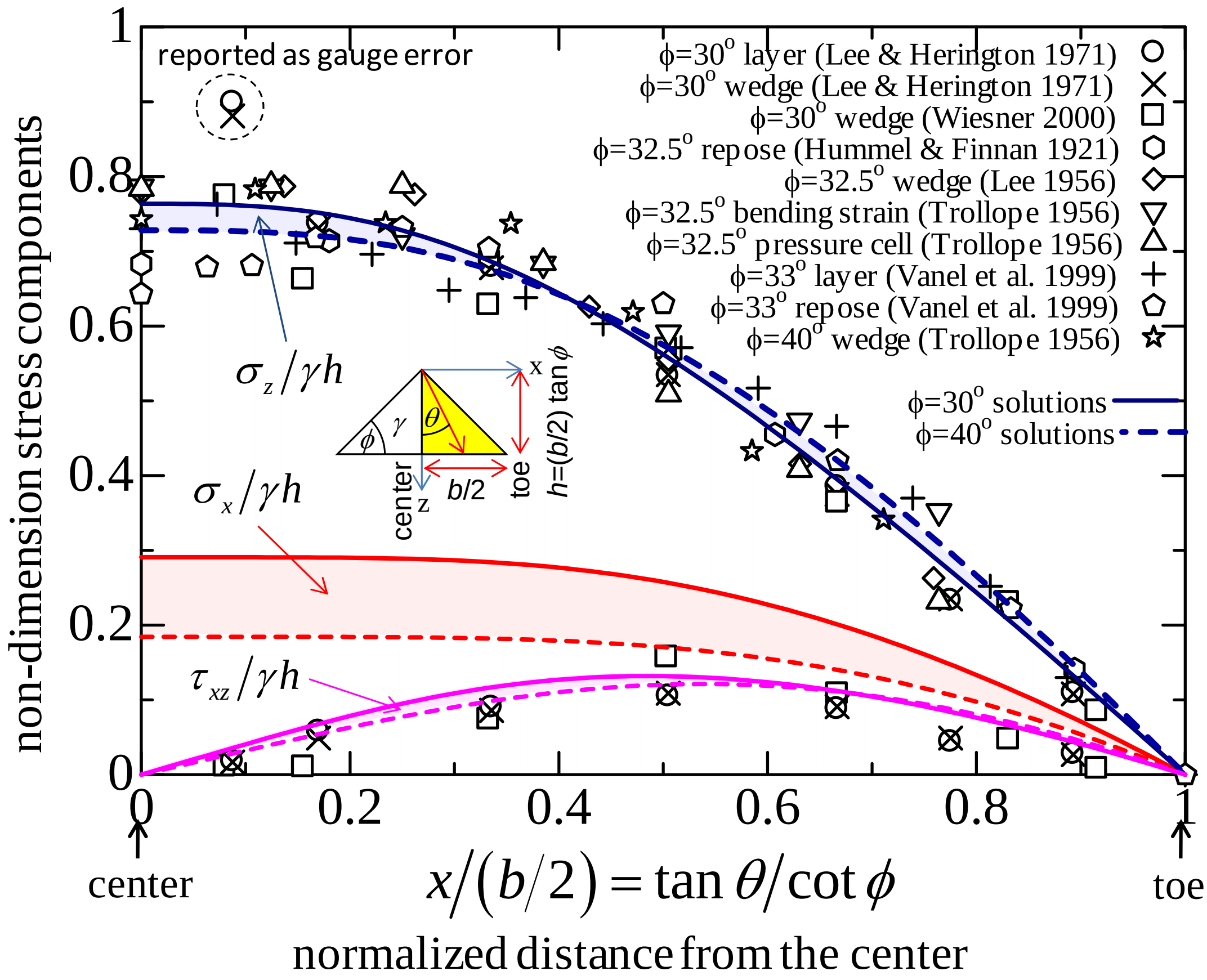

Figure 12 Normalized stress distribution along the normalized half-width base obtained from the theoretical and the experimental data. Symbols are experimental data obtained from past studies on wedge-shaped granular heaps carried out during 1921-2000 on various granular media with $\phi$=30°, 32.5°, 33° and 40°. Solid and dashed curves represent analytical solutions for $\phi$=30° and 40°, respectively, while shaded areas represent their in-between ranges.

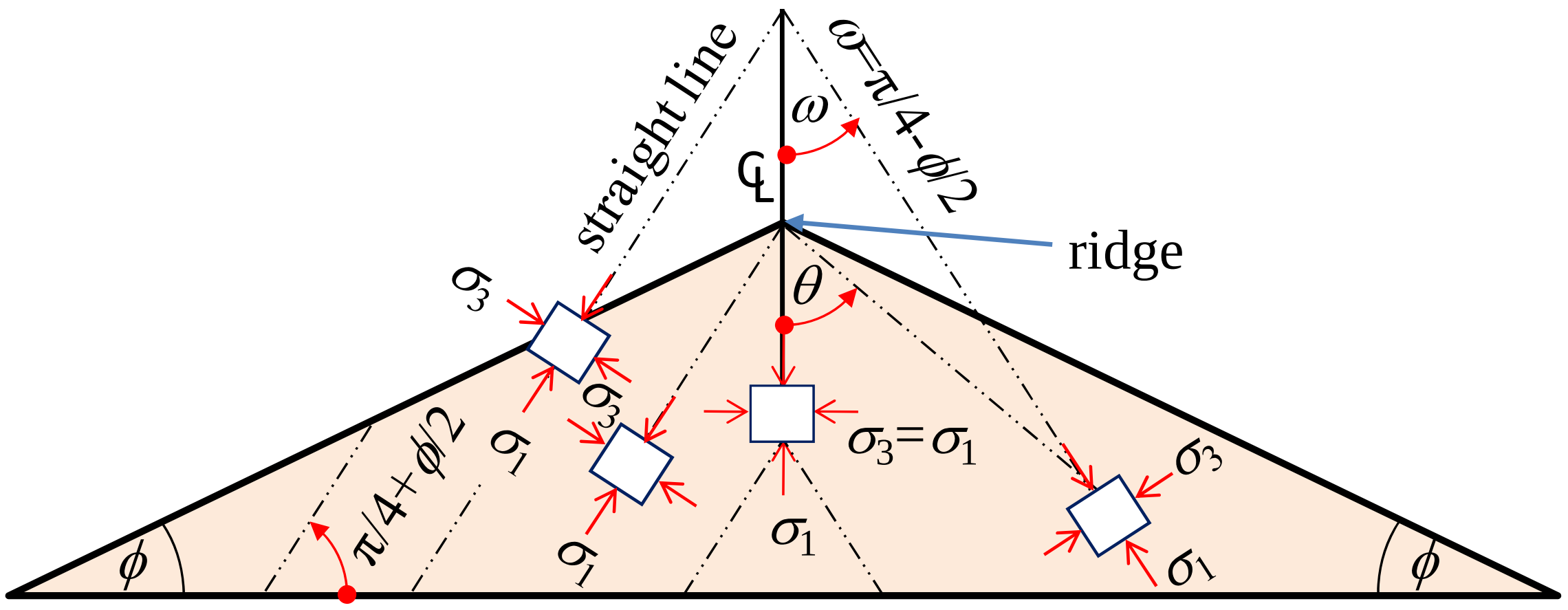

Figure 13 Schematization of stress orientation by the closure of fixed principal axes (FPA). The directions of major compression lie along fixed parallel straight lines, angled from the vertical in the middle between the sliding surface and the gravitating direction. The stress state along the centerline becomes isotropic.

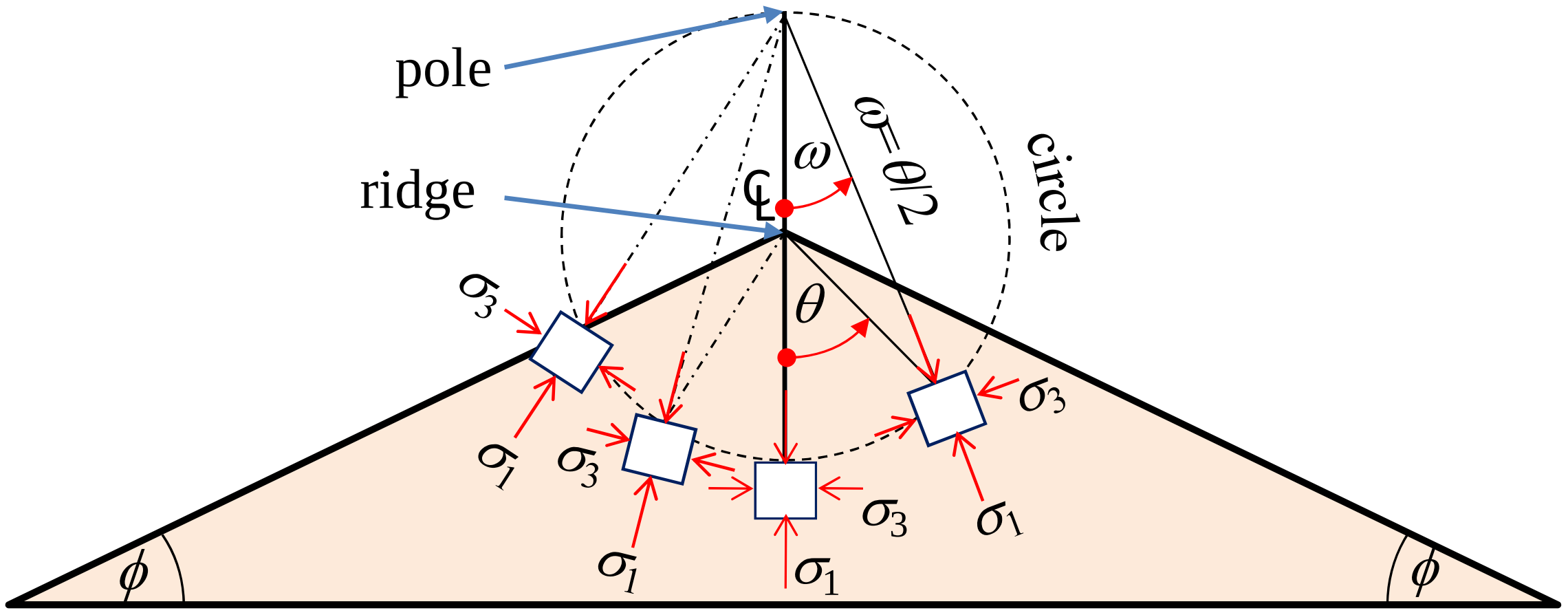

Figure 14 Schematization of stress orientation by the closure of polarized principal axes (PPA). The directions of major compression along a circle traced about the ridge point towards the pole of the circle. Any straight line drawn from the pole to the bulk will intersect the circle at angle equal to half of its central angle to the vertical.

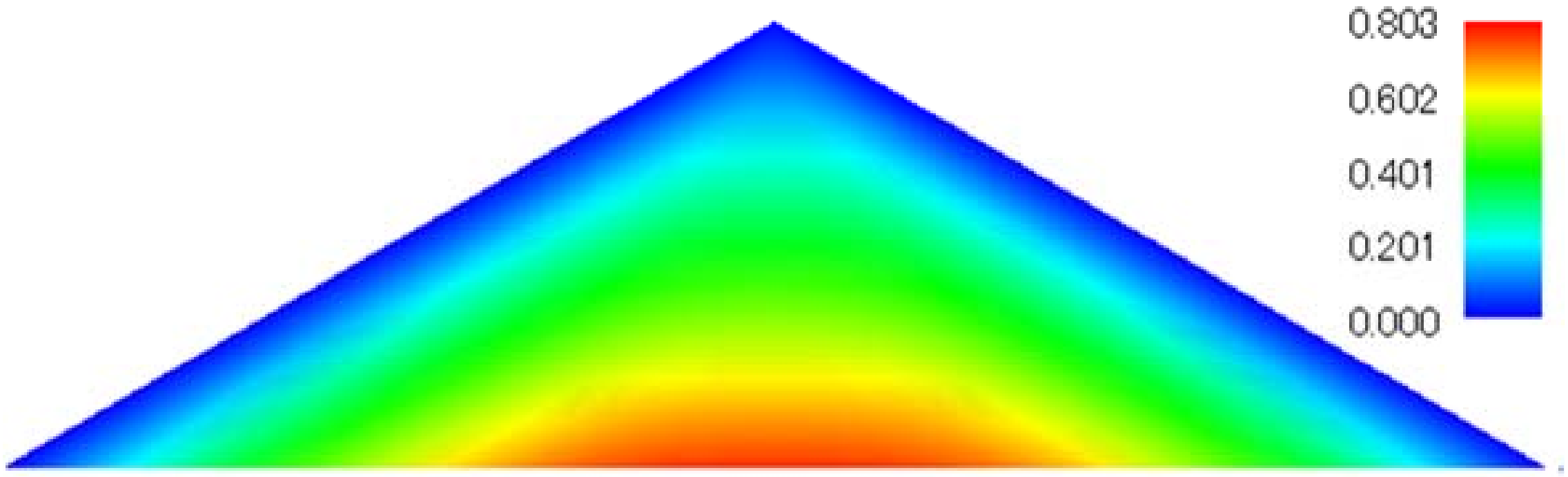

Figure 15 Contour of scaling vertical stress of sand heap ($\sigma_z/\gamma h$) for a given height $h$ and bulk unit weight $\gamma$ with angle of repose $\phi$=30$^o$.

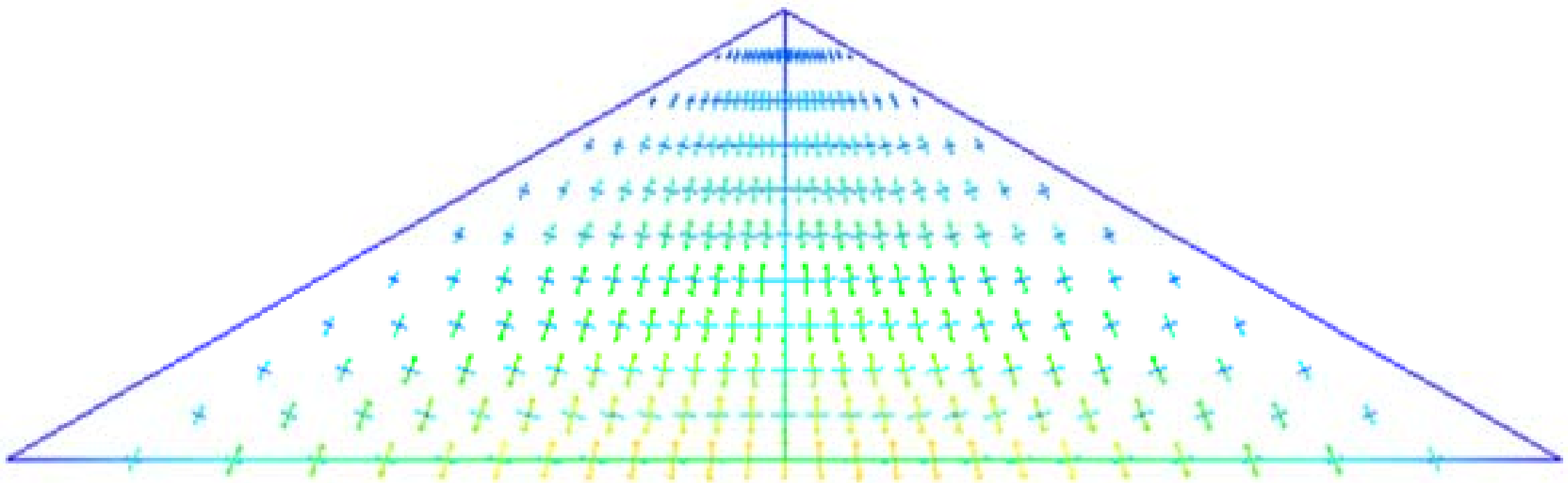

Figure 16 Variation of principal stress orientations in the bulk of a symmetrical sand heap with a triangular cross-section and $\phi$=30$^o$.

Table 1 Comparisons of various solutions for scaled stress variable $\chi$ and frictional variable $\beta$

| Solutions | $\chi$ | $\beta$ |
|---|---|---|
| (1) Nadai's original solution (1963) | $\dfrac{\cos\theta-\sin\phi}{\cos^2\phi}$ | $\sin\phi$ |
| (2) Nadai's alternative solution | $\dfrac{\cos\theta-\sin\phi}{\cos 2\phi+\sin\phi\cos\theta}$ | $\sin\phi$ |
| (3) Marais's corrected solution (1969) | $\dfrac{\cos\theta-\sin\phi}{\cos 2\phi+\dfrac{\sin^2\phi}{\cos\phi}\ln\left(\dfrac{1+\cos\phi}{\sin\phi}\right)}$ | $\sin\phi$ |
| (4) Exact solution | $\left(1-\left(\dfrac{\pi}{2}-\phi-\theta+\tan\theta\right)\tan\phi\right)\dfrac{\cos\theta}{\cos^2\phi}$ | $\dfrac{\left(\dfrac{\pi}{2}-\phi-\theta+\sin\theta\cos\theta\right)\tan\phi-\sin^2\phi}{2\left(\left(1-\left(\dfrac{\pi}{2}-\phi-\theta+\tan\theta\right)\tan\phi\right)\right)\cos\theta}$ |